\documentclass[%
 preprint, 
 amsmath,amssymb,
 aps, physrev,
]{revtex4-2}

\usepackage{graphicx}
\usepackage{dcolumn}
\usepackage{bm}
\usepackage{hyperref}
\hypersetup{
    colorlinks=true,
    linkcolor=blue,
    citecolor=blue,
    urlcolor=blue
}

\begin{document}

\preprint{}

\title{\textbf{The Effect of Corneal Topography and Mucins on Tear Film Rupture} 
}%

\author{Deepak Kumar}
\author{Pushpavanam S}%
 \email{Contact author: spush@iitm.ac.in}
\affiliation{Department of Chemical Engineering, Indian Institute of Technology Madras, Chennai, 600036, India}%

\date{\today}

\begin{abstract}
Tear film rupture on the corneal surface plays a critical role in ocular health and visual comfort. Conventional theoretical approaches often idealize the cornea as a perfectly smooth surface, ignoring the surface roughness that are characteristic of healthy as well as diseased eyes. In this study, we develop a comprehensive mathematical model to investigate tear film dynamics over the corneal surface incorporating the effects of surface roughness, slip, van der Waals forces, and lipid transport at the film-air interface. The corneal surface is represented by a small-amplitude periodic modulation. Steady-state solutions obtained using asymptotics reveal nonlinear corrections to the base profile at $O(\eta^2)$, which are confirmed numerically. Linear stability analysis performed using the Floquet theory demonstrates that an increase in the amplitude of roughness destabilizes the film. Specifically, both the dominant growth rate and the most unstable wavenumber increase with the roughness amplitude. Nonlinear simulations show that surface roughness significantly accelerates tear-film rupture. The slip coefficient, amplitude of roughness of the corneal surface and the initial film profile are found to significantly influence the rupture time. Moreover, the location of the rupture is sensitive to the initial disturbance. These results highlight the crucial role of surface topography and slip in determining tear film stability. The predicted rupture times are consistent with the experimental observations. The proposed model provides a realistic and accurate prediction of tear film dynamics and rupture over the corneal surface. This study offers a new perspective on tear film instability and will help address challenges such as contact lens failure  which is related to tear film behavior.
\clearpage
\end{abstract}

\maketitle



\section{Introduction}
\label{sec:Introduction}

\begin{figure}
  \centering
  \includegraphics[width=1\textwidth]{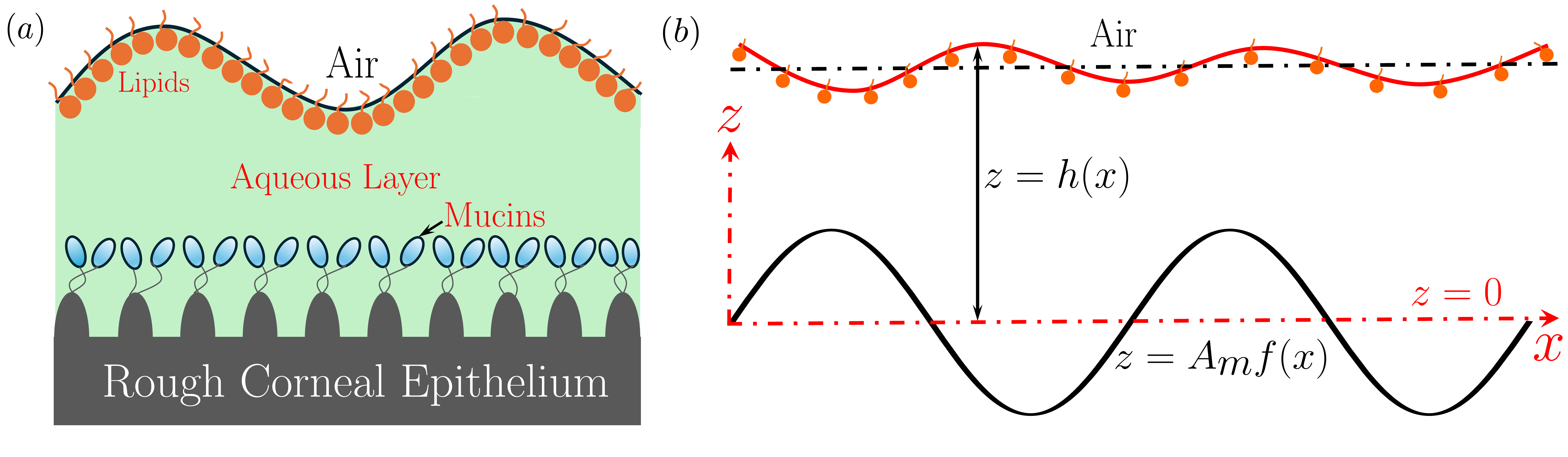}
  \caption{$(a)$ Illustration of the tear film over a rough corneal epithelial surface $(b)$ schematic representation of the mathematical model describing the dynamics of the tear film.}
  \label{fig:1}
\end{figure}

 The tear film is a thin fluid layer that coats the ocular surface and plays a vital role in maintaining optical clarity, delivering nutrients, and protecting the eye from environmental stress \citep{mantelli2008functions,selinger1979resistance}. Structurally, the tear film is comprised of three distinct layers, as illustrated in FIG \ref{fig:1}$(a)$. The outermost lipid layer, secreted by the Meibomian glands reduces evaporation and stabilizes the film by enhancing surface tension \citep{bron2004functional,craig1997importance,mcculley1997compositional,zhang2003analysis,zhang2003surfactant}. The middle layer is aqueous, which is produced primarily by the lacrimal glands. It forms the bulk of the tear film and supplies oxygen, nutrients, and antimicrobial agents to the cornea \citep{mcdermott2013antimicrobial}.  The innermost mucin layer is secreted by conjunctival goblet cells. This layer promotes uniform spreading of the tear film by reducing surface hydrophobicity and anchoring the film to the corneal epithelium surface \citep{cho1991stability,hodges2013tear}. The mucins are large glycoproteins expressed on the superficial surface of corneal epithelial cells \citep{davidson2004tear}. They form a hydrated, soft and flexible layer over the corneal epithelium. This facilitates partial movement of the tear film over the cornea, inducing a  partial slip \citep{braun2007model}. Specifically, the traditional no-slip boundary condition is not appropriate at the corneal surface. The slip coefficient depends on the physiological condition of the eye and may vary significantly between a healthy eye and an infected eye. An unstable tear film may result in dry eye syndrome and other ocular surface disorders \citep{willcox2017tfos}. A clear understanding of tear film dynamics is therefore crucial for improving diagnosis, developing therapies, designing ophthalmic devices and ocular drug delivery systems.\\
 Tear film breakup is influenced by several mechanisms, including van der Waals forces \citep{craster2009dynamics,sharma1985mechanism}, evaporation \citep{braun2018tear,peng2014evaporation} and nonuniform lipid distribution \citep{siddique2015tear,zhang2003analysis,zhang2003surfactant}. Much of the current understanding of tear film rupture originates from single-layer models \citep{dey2019model,jones2005dynamics,siddique2015tear,craster2009dynamics}. These studies typically assume an initially uniform and flat tear film over a smooth corneal surface. They demonstrate the significance of nonhydrodynamic forces, particularly van der Waals interactions in driving tear film breakup. These forces are distance dependent interactions that become significant at micro to nanometer scales. These forces are relevant in the context of tear film dynamics, as here the film  thickness is of the order of a few micro-meters. While tear film instability is primarily driven by van der Waals forces, the interfacial tension and Marangoni stresses arising from the surface lipids (which act as insoluble surfactants) have a stabilizing influence \citep{berger1974surface,zhang2004rupture}. These studies show that while capillary and Marangoni forces stabilize the film, van der Waals forces destabilize it. The Hamaker constant determines the strength of van der Waals interactions, and higher values promote faster rupture \citep{dey2019model}. Furthermore, increasing the slip coefficient at the corneal surface has been shown to accelerate film instability \citep{zhang2003analysis}. \\
 The thickness of the tear film is much lower than the radius of the cornea. This allows a simplification of the governing equations for tear film dynamics using the thin-film approximation \citep{de1994nonlinear,zhang2004rupture,zhong2019mathematical}. This yields a reduced-order model that captures the essential features of film dynamics and helps estimate the location and and time of rupture. Several studies have applied the lubrication approximation to thin films on rough or patterned substrates \citep{lin2010thin,vellingiri2015absolute,wierschem2003instability}. However, most of these works focus on gravitational or wettability-driven flows. They do not account for the physiological complexity of tear film dynamics, such as the presence of lipids and mucins \citep{kargupta2002dewetting,kondic2002flow,kondic2003flow}. \\
 
 Linear stability analysis has been used extensively to study tear film stability in both single and two-layer models \citep{dey2019model,zhang2003surfactant}. In these studies, the corneal surface is assumed to be perfectly smooth which leads to a base state with a uniform film thickness. Consequently, the linearized equations have constant coefficients and classical normal-mode analysis is applied. However, normal-mode analysis is no longer applicable when the linearized equations contain spatially periodic coefficients. In such cases, Floquet theory provides the appropriate framework for linear stability analysis \citep{kuchment2012floquet,pettas2022stability,ajaev2016stability}. Ajaev \textit{et al} performed a stability analysis using Floquet theory for a thin film flowing over gas-filled grooves \citep{ajaev2013application}. In this study, the slip length varies spatially along the grooves, leading to periodic coefficients in the linearized evolution equations, although the base state remains spatially uniform. Discretized eigenvalue method is an alternative and more general numerical approach. In this method, the perturbation field and the spatially periodic coefficients are both expanded as Fourier series over a large computational domain containing many periods of the substrate pattern \citep{ajaev2013application,jutley2018stability}.
 
 Gipson \textit{et al.} presented an electron micrograph of the tear film surface over the corneal epithelium highlighting the inherent roughness of the epithelial surface (\citep{gipson2003role}, Figure 4).  This observation is further supported by tear film interferometry, which reveals spatial variations associated with corneal surface irregularities \citep{king2014tear}. The tear film fills microscopic irregularities in the corneal epithelium creating a smooth and continuous optical surface. The standard deviation of corneal surface height is approximately 0.129 $\mu m$ \citep{king2014tear}. Previous models of tear film breakup have typically assumed a smooth corneal surface. This leads to a symmetric evolution of the film. The effect of  roughness of the corneal surface on  the dynamics of the tear film has not been explored. Pathological conditions such as epithelial cell breakdown, stromal remodeling, and inflammatory infiltration further alter corneal topography \citep{kamil2021corneal,lee2016ins,ljubimov2015progress}. Both the surface roughness and slip coefficient vary substantially under disease conditions \citep{liu1999corneal,mccafferty2012corneal}. For instance, epithelial ulcers increase surface roughness and disrupt the smooth epithelium. Hence, a comprehensive model that accounts for surface roughness with partial slip is necessary to accurately describe tear film dynamics and its rupture under realistic ocular conditions. 

In this work, we consider the limit where the amplitude of corneal surface roughness is small compared with the mean film thickness. We derive solutions that capture both smooth-surface dynamics and roughness-induced corrections. Linear stability analysis is carried out using Floquet theory to determine the most unstable wavelength of perturbations. This is also verified using the discretized eigenvalue method. The nonlinear numerical simulations are performed to investigate the spatio-temporal evolution of the film. The governing equations are solved using a Fourier spectral method, which provides high accuracy and efficiency for periodic domains \citep{guo2013application,trefethen2000spectral,weideman2000matlab}. The simulations capture the full nonlinear evolution of the tear film including thinning and rupture.\\

 The paper is organized as follows. Section \ref{sec:Mathematical Formulation} presents the formulation of the governing equations based on the lubrication approximation. A linear stability analysis is then performed in the rough domain in section \ref{sec:LSA}. This is followed by nonlinear simulations to investigate the effects of the slip coefficient, roughness amplitude, and initial disturbance on tear film rupture in section \ref{sec:nonlin}. Section \ref{sec:conclusions} summarizes the key findings, highlights the physiological implications, and discusses possible extensions of the present model.
\section{Mathematical formulation}
\label{sec:Mathematical Formulation}

\subsection {Problem description}
 Tear film interferometry has revealed spatial variations in the corneal surface. This indicates that it is inherently a rough surface \citep{king2014tear}. The tear film fluid fills microscopic irregularities in the corneal epithelium creating a smooth and continuous optical surface \citep{DOUGHTY20021}. In this work, we consider a thin tear film spread over a rough corneal surface represented as $z=A_m  f(x)$, where $A_m$ denotes the amplitude of surface roughness in $z$-direction and $f(x)$ describes the spatial variation along the $x$-direction as shown in FIG \ref{fig:1}$(b)$. Since the thickness of the tear film is much smaller than the radius of the cornea \citep{braun2012dynamics,dey2019model,jones2005dynamics,zhang2003analysis}, the governing equations and boundary conditions are simplified using the lubrication approximation.
The lipids present on the free surface are modelled as insoluble solutes that advect and diffuse along the interface. The surface tension denoted by $\sigma$ depends on the local lipid concentration $(\gamma)$. As a first step, we assume isothermal conditions and neglect evaporation from the free surface and loss of fluids due to osmosis from the ocular surface. The  mucin layer on the corneal surface reduce friction by providing slip over the corneal surface. The mucin layer is incorporated by applying a partial slip boundary condition at the corneal surface with the slip length/coefficient $(\beta)$.

The thin tear film occupies the region bounded by the corneal surface defined as $z = A_m f(x)$ and the tear film-air interface denoted by $z = h(t,x)$. The fluid dynamics within this region are governed by the continuity equation and the Navier-Stokes equations. In the two-dimensional Cartesian coordinate system illustrated in FIG \ref{fig:1}$(b)$, these governing equations are:
\begin{equation}
\nabla \cdot \boldsymbol{v} =0,
  \label{eq1}
\end{equation}
\begin{equation}
\rho \left( \frac{\partial \boldsymbol{v}}{\partial t} 
+ \boldsymbol{v} \cdot \nabla \boldsymbol{v} \right)
= - \nabla (p + \phi) + \mu \nabla^{2} \boldsymbol{v}.
  \label{eq2}
\end{equation}
where $\boldsymbol{v}$= $(u,w)$ is the velocity vector and $p$ is the pressure.  $u$ and $w$ represent the velocity components in the $x$ and $z$ directions respectively. The thin film flow is driven by a pressure gradient arising from non-uniform van der Waals forces, represented by the potential $\phi$.  The value of $\phi$ depends upon the thickness of the tear film above the corneal surface i.e. $\phi=\frac{A_k}{(h(t,x)-A_m  f(x))^3}$,  where $A_k$ is a Hamaker constant.\\
Additionally, the lipid concentration $\gamma(t,x)$ is governed by the transport equation at the surface of the tear film at $z=h(t,x)$ as,
\begin{equation}
\frac{\partial \gamma}{\partial t}
+ \nabla_{s} \cdot (\gamma \boldsymbol{v}_{s})
+ \gamma (\nabla_{s} \cdot \boldsymbol{n})(\boldsymbol{v} \cdot \boldsymbol{n})
= D_{s} \nabla_{s}^{2} \gamma .
\label{eq3}
\end{equation}
where $\boldsymbol{n}$ is the normal vector to the free surface $z=h(t,x)$ and $\nabla_{s}=(\boldsymbol{I}-\boldsymbol{nn}) \cdot \nabla$ is the surface divergence operator. The surface velocity vector is given by $\boldsymbol{v}_s=\boldsymbol{v}-\boldsymbol{nn} \cdot \boldsymbol{v}$  and $D_s$ represents the surface diffusivity of the insoluble lipids. We assume the lipid layer is dilute, and the interfacial tension $\sigma$ decreases linearly with the lipid concentration $\gamma$, $\sigma(\gamma)=\sigma_m-S \gamma/\gamma_m$ \citep{zhang2003surfactant}. Here, $\sigma_m$ is the maximal interfacial tension on the lipid-free interface, $S$ is the maximal spreading pressure and $\gamma_m$ is the maximum lipid concentration.\\
On the corneal surface $z=A_m  f(x)$, we impose the partial slip boundary condition with a slip length $\beta$ together with a no-penetration condition. Hence,
\begin{equation}
\boldsymbol{v}_{t} = \beta \, (\boldsymbol{n}_{c} \cdot \nabla \boldsymbol{v}_{t}),
\qquad
\boldsymbol{n}_{c} \cdot \boldsymbol{v} = 0 .
\label{eq4}
\end{equation}
where, $\boldsymbol{v}_t=\boldsymbol{t}_c \cdot \boldsymbol{v}$ denotes the tangential velocity along the corneal surface. Here, $\boldsymbol{t}_c$ and $\boldsymbol{n}_c$ are the unit tangent and normal vectors respectively defined on the corneal surface $z=A_m  f(x)$. The expressions for these vectors are provided in the supplementary material (Section S.1).\\
The tangential stress balance condition at the free surface is given as,
\begin{equation}
\boldsymbol{n} \cdot \boldsymbol{\tau} \cdot \boldsymbol{t} = \nabla_s \sigma \cdot \boldsymbol{t}
\label{eq5}
\end{equation}
Here, $\boldsymbol{\tau}$ is the deviatoric stress tensor and $\boldsymbol{t}$ is tangent vector on the surface $z=h(t,x)$. The normal stress balance gives the relationship between surface tension and pressure jump at the free surface $z=h(t,x)$. 
\begin{equation}
\boldsymbol{n} \cdot \boldsymbol{\tau} \cdot \boldsymbol{n} =(p-p_\text{atm})-\sigma(\nabla \cdot \boldsymbol{n})
\label{eq6}
\end{equation}
where, $p$ denotes the pressure within the liquid film and $p_\text{atm}$ is the pressure above the tear-air interface. The expressions for these terms have been given in the supplementary material (Section S.1). The surface of the tear film $z=h(t,x)$ evolves according to the kinematic boundary condition,
\begin{equation}
\frac{\partial h}{\partial t}
+ u \frac{\partial h}{\partial x}
= w .
\label{eq7}
\end{equation}
The physical parameters of the tear film and the corneal surface used in this study  are summarized in TABLE \ref{tab:1}. 
\subsection {Nondimensionalization}
The governing equations and the boundary conditions are nondimensionalized using the following characteristic scales,\\
$$x_c =L, z_c=H,u_c=U=\frac{A_k}{6 \pi \mu H L},w_c =\epsilon U, t_c = \frac{L}{U}$$
$$p_c=\frac{A_k}{6 \pi H^3},h_c=H,\beta_c=H,\gamma_c=\gamma_m,\sigma_c=\sigma_m$$
Here, $L$ and $H$ represent the characteristic length scales in $x$ and $z$ directions respectively.  Since van der Waals force is the dominant mechanism driving tear film rupture, the characteristic velocity scale $U$ is determined by balancing viscous forces with van der Waals forces.
\begin{table}
\begin{center}
\def~{\hphantom{0}}
\begin{tabular}{lll}
\textbf{Symbol} & \textbf{Description} & \textbf{Value (Reference)} \\[5pt]
$H$ & Characteristic thickness & $0.6\times10^{-6}\ \mathrm{m}$ \citep{dey2019model} \\
$L$ & Characteristic length & $1.5\times10^{-4}\ \mathrm{m}$ \citep{luke2021parameter} \\
$\rho$ & Density of tear film & $1000\ \mathrm{kg\,m^{-3}}$ \citep{deng2014heat} \\
$\mu$ & Viscosity of tear film & $1.3\times10^{-3}\ \mathrm{Pa\,s}$ \citep{tiffany1991viscosity} \\
$\sigma_m$ & Maximum interfacial tension & $4.5\times10^{-2}\ \mathrm{N\,m^{-1}}$ \citep{nagyova1999components} \\
$\beta$ & Slip coefficient & $3.5\times10^{-7}\ \mathrm{N\,m^{-1}}$ \citep{zhang2003analysis} \\
$S$ & Maximum spreading pressure & $7.5\times10^{-8}\ \mathrm{N\,m^{-1}}$ \citep{zhang2003analysis} \\
$\gamma_m$ & Maximum lipid concentration & $4\times10^{-7}\ \mathrm{mol\,m^{-2}}$ \citep{bruna2014influence} \\
$A_k$ & Unretarded Hamaker constant & $6 \pi \times 3.5\times10^{-19}\ \mathrm{Pa\,m^{3}}$ \citep{winter2010model} \\
$D_s$ & Surface diffusivity & $10^{-11}\ \mathrm{m^{2}\,s^{-1}}$ \citep{adalsteinsson2000lipid} \\
\end{tabular}
\caption{Physical parameters used in the mathematical model and their corresponding reference sources.}
\label{tab:1}
\end{center}
\end{table}
\subsection {Lubrication approximation (rough corneal surface)}
We define $ \epsilon =\frac{H}{L} \ll 1 $ and exploiting this, we apply the lubrication approximation to our system. Under this framework, only leading-order terms are retained, while terms of order $\epsilon$ or smaller are neglected. The resulting nondimensionalized governing equations are derived in the Supplementary Material (Section S.2). The corneal surface is defined as $z=\eta f(x)$ in dimensionless form, where $\eta=\frac{A_m}{H}$ denotes the ratio of the surface roughness amplitude to the characteristic film thickness. The dimensionless variables with superscript  $(*)$ in the Supplementary material are now used without superscript. As our goal is to investigate tear-film dynamics over a realistic rough corneal surface, we approximate the corneal roughness $f(x)$ using a sinusoidal profile in this study. This choice captures the essential roughness in the human cornea and is consistent with the experimental surface topography reported by Gipson and Argueso (\cite{gipson2003role}, Figure 4). This simplification not only makes the problem tractable but also preserves the key physical features of the tear film behavior on the corneal surface. Accordingly, we define $f(x)$ in the domain $[0,1]$ as,\\
\begin{equation}
f(x)= \sin(kx),
\label{eq8p}
\end{equation}
As the corneal roughness is defined as $z=\eta f(x)$, the surface is continuous and differentiable across the entire domain. In healthy eyes, the standard deviation of corneal surface roughness is typically reported as $0.129$ $\mu m$ \citep{gipson2003role}. To obtain this characteristic roughness, we set $\eta=0.26$ which yields a dimensional standard deviation of $0.129$ $\mu m$ for the corneal roughness given by equation (\ref{eq8p}). This value is used in the simulations to represent the small-amplitude roughness characteristic of a normal corneal surface. The characteristic wavelength of corneal roughness is not well established in the literature. In the absence of definitive measurements, we take $k=2 \pi$, corresponding to a single spatial period over the computational domain. However, we show later (in FIG {\ref{fig:5}} $(a)$) that the stability characteristics are weakly sensitive to the choice of frequency of the corneal roughness.
At leading order, the lubrication approximation yields
\begin{equation}
\frac{\partial u}{\partial x}+\frac{\partial w}{\partial z}=0,
\label{eq8}
\end{equation}
\begin{equation}
-\frac{\partial p}{\partial x}-\frac{\partial \phi}{\partial x} + \frac{\partial^2 u}{\partial z^2}=0,
\label{eq9}
\end{equation}
\begin{equation}
-\frac{\partial p}{\partial z}=0.
\label{eq10}
\end{equation}
 The dimensionless  lipid transport equation (\ref{eq3}) in dimensionless form is given as,
\begin{equation}
\frac{\partial \gamma}{\partial t}
+ \frac{\partial}{\partial x}\!\left( u_{s} \gamma \right)
= \frac{1}{\mathrm{Pe}_{s}}
\frac{\partial^{2} \gamma}{\partial x^{2}} .
\label{eq11}
\end{equation}
Here, $u_s$ denotes the $x$-component of surface velocity and $Pe_{s}=UL/D_s$  is the Peclet number for mucin diffusion. This is subject to  the boundary conditions,\\
At $z=\eta f(x)$, 
\begin{equation}
u=\beta \frac{\partial u}{\partial z},                        
\label{eq12}
\end{equation}
and 
\begin{equation}
w-\eta f'(x) u=0.                        
\label{eq13}
\end{equation}
At $z=h(x)$, 
\begin{equation}
p=-C \frac{\partial^2 h}{\partial x^2},                        
\label{eq14}
\end{equation}
and 
\begin{equation}
\frac{\partial u}{\partial z} =-M \frac{\partial \gamma}{\partial x} .  \label{eq15}
\end{equation}
Here,  $C=\frac{\epsilon^3 \sigma_m}{\mu U}$ is the reduced Capillary number and $M= \frac{SH}{\mu U L }$ is the Marangoni number. Integrating the  continuity equation in the $z$-direction from $\eta f(x)$ to  $h(t,x)$ , we obtain\\
\begin{equation}
\int_{\eta f(x)}^{h(t,x)} 
\frac{\partial u}{\partial x}\, \mathrm{d}z
+ \left. w \right|_{z = h(t,x)}
- \left. w \right|_{z = \eta f(x)}
= 0 .  
\label{eq16}
\end{equation}
Using Leibniz's rule of integration and the kinematic boundary condition yields,
\begin{equation}
\frac{\partial h}{\partial t}
+ \frac{\partial}{\partial x}
  \int_{\eta f(x)}^{h(t,x)} u\, \mathrm{d}z
+ \eta f'(x)\,
  \left. u \right|_{z = \eta f(x)}
- \left. w \right|_{z = \eta f(x)}
= 0 .
\label{eq17}
\end{equation}
 The dependent variable $u$ and $w$ are functions of time and spatial coordinates $x$ and $z$, while $p,\phi,h$ and $\gamma$ depend only on time and coordinate $x$.\\
 Since the upper surface of the tear film is time dependent, the physical domain is   $z \in [\eta f(x),h(t,x)]$. To avoid solving the governing equations on a time dependent domain, we introduce a coordinate transformation that maps the deforming film region into a fixed rectangular domain. For this, we define a transformed vertical coordinate,
 \begin{equation}
\zeta = \frac{z-\eta f(x)}{h-\eta f(x)}.
\label{eq18}
\end{equation}
This transformation maps the physical tear-film region onto a fixed rectangular computational domain defined by $x\in[0,1]$ and $\zeta \in [0,1]$. All governing equations and boundary conditions are subsequently expressed in the $(x,\zeta)$ domain. The transformed continuity equation is 
\begin{equation}
\frac{\partial u}{\partial x}
+ \left(
\frac{-\,\eta f'(x)}{h - \eta f(x)}
- \frac{\zeta \left( h' - \eta f'(x) \right)}{h - \eta f(x)}
\right)
\frac{\partial u}{\partial \zeta}
+ \frac{1}{h - \eta f(x)} \,
\frac{\partial w}{\partial \zeta}
= 0 .
\label{eq19}
\end{equation}
The primes $(')$ denote the derivative with respect to $x$. The curvature of the tear film governs the pressure distribution within the film, as given by equation (\ref{eq14}). Using this, equation (\ref{eq9}) reduces to 
\begin{equation}
C \frac{\partial^3 h}{\partial x^3} +\frac{3A_k(h'-\eta f'(x))}{(h-\eta f(x))^4}+\frac{1}{(h-\eta f(x))^2} \frac{\partial^2 u}{\partial \zeta^2}=0.
\label{eq20}
\end{equation}
The surfactant transport equation and kinematic boundary conditions remain unchanged, as they are defined at $\zeta=1$. The boundary conditions (\ref{eq12}), (\ref{eq13}) and (\ref{eq15}) are,
\begin{equation}
u=\frac{\beta}{h-\eta f(x)} \frac{\partial u}{\partial \zeta}  \quad \text{at} \quad \zeta =0,                      
\label{eq21}
\end{equation}
\begin{equation}
w-\eta f'(x) u=0    \quad \text{at} \quad \zeta =0,                    
\label{eq22}
\end{equation}
and 
\begin{equation}
\frac{1}{h-\eta f(x)}\frac{\partial u}{\partial \zeta} =-M \frac{\partial \gamma}{\partial x} \quad \text{at} \quad \zeta =1.
\label{eq23}
\end{equation}
Integrating equation (\ref{eq20}) twice with respect to $\zeta$ yields:
\begin{equation}
u=c_1 (x)  \frac{\zeta^2}{2}+c_2 (x)  \zeta+c_3 (x),
\label{eq24}
\end{equation}
where, $c_1(x)=- (h - \eta f(x))^2\left(\frac{3 A_k \left( h' - \eta f'(x) \right)}{(h - \eta f(x))^4}+ C h'''\right)$. Using equation (\ref{eq21}), $c_3=\frac{\beta c_2 (x)}{h-\eta f(x) }$. The tangential stress balance equation (\ref{eq23}) implies  
\begin{equation}
c_2 (x)=M\gamma' (\eta f(x)-h)-c_1 (x).
\label{eq25}
\end{equation}
Substituting the values of $c_1 (x)$ and $c_2 (x)$ in equation (\ref{eq24}) yields,
\vspace{-1em}
\begin{equation}
\begin{aligned}
u =\Bigg(
M \gamma' (\eta f(x) - h)
+ (h - \eta f(x))^2 
\left(
\frac{3 A_k \left( h' - \eta f'(x) \right)}{(h - \eta f(x))^4}
+ C h'''
\right)
\Bigg) 
\left(
\frac{\beta}{h - \eta f(x)} + \zeta
\right) & \\
 - (h - \eta f(x))^2 
\left(
\frac{3 A_k \left( h' - \eta f'(x) \right)}{(h - \eta f(x))^4}
+ C h'''
\right)
\frac{\zeta^2}{2} 
\end{aligned}
\label{eq26}
\end{equation}
From continuity equation (\ref{eq19}) and equation (\ref{eq24}), we obtain 
\begin{equation}
\begin{aligned}
w ={}& \frac{\zeta^3}{6}
\left(
 c_1'(x)(\eta f(x) - h) 
+ 2c_1(x)(-\eta f'(x)
+   h')
\right) \\
&+ \frac{\zeta^2}{2}
\left(
 c_2'(x)(\eta f(x) - h )
+\eta f'(x)( c_1(x) 
-  c_2(x))
+ c_2(x) h'
\right) \\
&+ \zeta
\left(
c_3'(x)(\eta f(x) 
-  h )
+  \eta c_2(x) f'(x)
\right) + \eta c_3(x) f'(x)
\end{aligned}
\label{eq27}
\end{equation}

Substituting the expression of $u$ into equation (\ref{eq11}) and equation (\ref{eq17}), we obtain the evolution equations for film thickness and the lipid concentration, 
\begin{equation}
\begin{aligned}
\frac{\partial h}{\partial t}
&={} \frac{A_k}{(h-\eta f(x))^3}
\Bigg(
\eta h^2 f''(x)
+ 3\beta \eta h f''(x)
- 2\eta^2 h f(x) f''(x)
- 3\beta \eta^2 f(x) f''(x) 
\\
& + \eta^3 f(x)^2 f''(x)
+ 2\eta h' f'(x)(-6\beta+\eta f(x)-h)
+ \eta^2 f'(x)^2 (6\beta-\eta f(x)+h) \\
&
+ h''(\eta f(x)-h)(3\beta-\eta f(x)+h)
- \eta f(x)h'^2
+ 6\beta h'^2
+ h h'^2
\Bigg) \\
&- \frac{C}{3}(-h+\eta f(x))
\Bigg(
(h+3\beta-\eta f(x))(-h+\eta f(x)) h'''' 
+ 3(h+2\beta-\eta f(x)) h''' (-h'+\eta f'(x))
\Bigg) \\
&- M \Bigg(
\gamma'(\beta-\eta f(x)+h)(\eta f'(x)-h')
- \frac{1}{2} \gamma'' (h-\eta f(x))(2\beta-\eta f(x)+h)
\Bigg),
\end{aligned}
\label{eq28}
\end{equation}
and,
\begin{equation}
\begin{aligned}
\frac{\partial \gamma}{\partial t}
={}& \frac{A_k}{2 (h-\eta f(x))^4}
\Bigg(
-3 h'
\Big(
4\gamma \eta f'(x)(3\beta-\eta f(x)+h)  
+ \gamma' (h-\eta f(x))(2\beta-\eta f(x)+h)
\Big)  \\
&\quad
+ 3\eta
\Big(
\gamma f''(x)(h-\eta f(x))(2\beta-\eta f(x)+h)  
+ \gamma' f'(x)(h-\eta f(x))(2\beta-\eta f(x)+h)  \\
&\qquad
+ 2\gamma \eta f'(x)^2 (3\beta-\eta f(x)+h)
\Big)  - 3\gamma h'' (h-\eta f(x))(2\beta-\eta f(x)+h)  \\
&\quad
+ 6\gamma h'^2 (3\beta-\eta f(x)+h)
\Bigg)  + C \Bigg(
\gamma \eta h''' f'(x)(\beta-\eta f(x)+h)  \\
&\qquad
- \frac{1}{2} (\gamma h'''' + h''' \gamma')
(h-\eta f(x))(2\beta-\eta f(x)+h) 
- \gamma h''' h' (\beta-\eta f(x)+h)
\Bigg)  \\
&+ M
\Big(
- \gamma \eta \gamma' f'(x)
+ (\gamma \gamma'' + \gamma'^2)(\beta-\eta f(x)+h)
+ \gamma \gamma' h'
\Big) + \frac{1}{Pe_s}\,\gamma'' .
\end{aligned}
\label{eq29}
\end{equation}

Equations (\ref{eq28}-\ref{eq29}) govern the nonlinear evolution of the film thickness and surfactant concentration over a rough corneal surface defined by $z=\eta f(x)$. In the limiting case $\eta=0$, the corneal surface becomes smooth, and the governing equations reduce to the classical evolution equations for $h$ and $\gamma$. In this limit, our formulation recovers the model of Zhang \textit{et al} \cite{zhang2003surfactant}, thereby demonstrating the mathematical consistency of the derivation. We first compute the steady-state solution, then examine its stability, and finally perform nonlinear simulations to characterize the spatiotemporal evolution of the tear film over the corneal surface.

\subsection{Steady state}
At steady state, the tear film is quiescent and the velocity field vanishes hence, $u=w=0$. The tangential stress balance (equation (\ref{eq23})) then requires the base-state lipid concentration to be spatially uniform ($\gamma=\gamma_s$). Consequently, surface-tension gradients are absent, and hence the Marangoni stress vanishes in the steady state configuration. The steady-state thickness profile $h(x)$ is governed by the balance between the Capillary forces and the Van der Waals forces in the $x$- momentum balance equation,
\begin{equation}
    Ch'''(x) + \frac{3 A_k(h'(x)-\eta f'(x))}{(h(x)-\eta f(x))^4}=0.
\label{eq30}
\end{equation}
This nonlinear third-order ordinary differential equation defines $h(x)$. The equation reflects a local balance between capillary pressure gradients and disjoining pressure induced by intermolecular forces. We define another variable $y(x)=h(x)-\eta f(x)$ which represents the effective thickness of the tear film over the corneal surface. Substituting this in equation (\ref{eq30}), we obtain,
\begin{equation}
    C\frac{d^3y}{dx^3} + \frac{3 A_k}{y^4} \frac{dy}{dx}=C\eta k^3 \cos(k x).
\label{eq31}
\end{equation}
The corneal roughness introduces a spatially periodic forcing into the governing equation through the term $ C \eta k^3 \cos(kx)$. The resulting problem is thus a forced third-order nonlinear ordinary differential equation for the steady film thickness.

\begin{figure}
  \centering
  \includegraphics[width=1\textwidth]{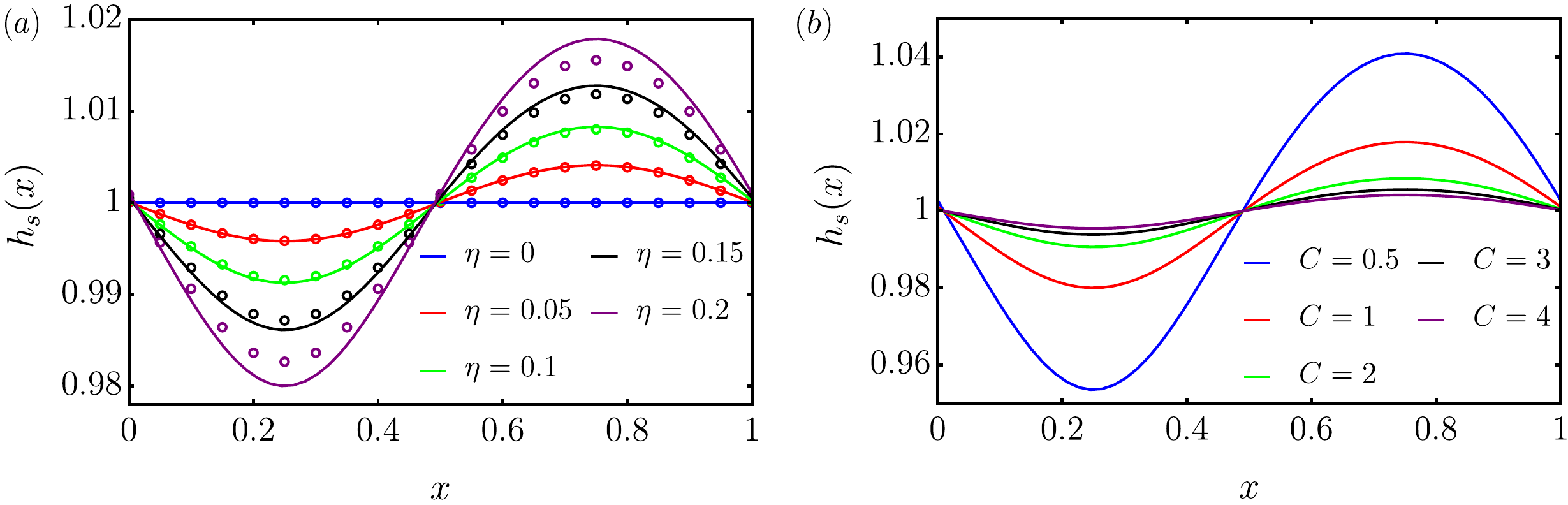}
  \caption{ Steady-state film profiles. $(a)$ Steady states for different values of $\eta$ with $A_k=1,C=1$ $(b)$  Steady states for different values of $C$ with $A_k=1,\eta=0.2$. Solid lines denote numerical solutions, while symbols indicate the corresponding asymptotic solutions.}
  \label{fig:2}
\end{figure}

\subsubsection{Asymptotic analysis in the weak-roughness limit}
Although the governing equation is nonlinear, an analytical solution can be obtained in the limit of weak surface roughness, $\eta \ll 1$. For this, the film thickness $y(x)$ can be expanded as a regular perturbation series,
\begin{equation}
   y(x) = y_0(x) + \eta y_1(x) +\eta^2 y_2(x) +...
\label{eq32}
\end{equation}
Here, $y_0 (x)$ represents the uniform base thickness for a smooth surface and higher-order terms describe the modulation induced by the underlying surface roughness. At the leading order, $ O(1)$, the governing equation reduces to
\begin{equation}
 Cy_0'''(x) + \frac{3 A_k}{y_0^4} y_0'(x)=0
\label{eq33}
\end{equation}
Subject to periodic boundary conditions, this equation admits only spatially uniform solutions, yielding $y_0 (x)$=constant. The detailed solution is provided in the Supplementary Material (Section S.3). This leading-order solution corresponds to a tear film over a smooth corneal surface $(\eta=0)$, for which the steady-state thickness is uniform. Without loss of generality, we normalize the base-state thickness and set $y_0 =h_{ss} =1$.
At $O(\eta)$, the governing equation becomes
\begin{equation}
 Cy_1'''(x) + \frac{3 A_k}{h_{ss}^4} y_1'(x)=C k^3 \cos(k x)
\label{eq34}
\end{equation}
This describes the linear response of weak surface roughness to the film thickness. Imposing periodic boundary conditions, the first-order correction is obtained as $y_1(x) = \frac{C k^2} {\left(\dfrac{3A_k}{h_{ss}^4} - C k^2 \right)}\, \sin(kx)$. At this order, the surface roughness induces a sinusoidal modulation in the film thickness with  the same wavenumber as that of the substrate. However, the amplitude is determined by the competition between capillary forces and van der Waals interactions. Next, we obtain the solution to $ O(\eta^2 ), $ \\

At $O(\eta^2 )$,\\
\begin{equation}
C y_2^{\,'''}
+
\frac{3A_k}{h_{ss}^4}
\left(
y_2'
-
\frac{4 y_1 y_1'}{h_{ss}}
\right)=0
\label{eq35}
\end{equation}
The solution for $y_2(x)=\frac{3 A_k B^2}{h_{ss}^5 \left( \dfrac{3A_k}{h_{ss}^4} - 4 C k^2 \right)} \, \cos(2kx)$ where $B=\dfrac{C k^2}{\frac{3A_k}{h_{ss}^4}-4Ck^2}$. Reconstructing $ h(x)$, we obtain 
\begin{equation}
h_s(x)
=
1
+ \eta (1+B)\sin(kx)
- \eta^2
\frac{3 A_k B^2}
{h_{ss}^5 \left( \dfrac{3A_k}{h_{ss}^4} - 4 C k^2 \right)}
\cos(2kx).
\label{eq36}
\end{equation}
Here, $ h_s (x)$ represents spatially periodic film thickness at the steady state. At $O(\eta^2)$, nonlinear terms  generate higher-harmonic  contributions ($\cos(2kx)$) to the steady state. At this order, nonlinearity feeds back into the base state and modifies the steady film profile beyond a purely sinusoidal response. FIG \ref{fig:2}$(a)$ shows steady-state film thickness profiles for different values of the surface roughness amplitude $\eta$. The asymptotic solutions obtained analytically from equation (\ref{eq36}) are indicated by open circles in FIG \ref{fig:2}$(a)$. The steady state film thickness depends on the reduced Capillary number $(C)$, roughness amplitude $(\eta)$ and wave number of the corneal surface $(k)$. 

\subsubsection{Numerical computation of steady-state solutions}
When the surface roughness amplitude is no longer asymptotically small, the perturbation series solutions derived above is not valid. For large $\eta$,  steady-state solutions are computed numerically. The steady-state equation is discretized using a Fourier pseudospectral method \citep{trefethen2000spectral,weideman2000matlab}  on the periodic domain $x \in [0,1]$. The corneal surface roughness $f(x)$ and its derivative $ f' (x)$ are discretized using $N_p$ equispaced collocation points. The spatial derivatives are approximated using Fourier differentiation matrices constructed from the discrete Fourier transform, providing spectrally accurate representations of derivatives up to third order. \\
Let $ h_i$ denote the discrete approximation to the steady film thickness at the collocation points, collected in the vector $ h \in R^{N_p }$. The steady-state governing equation (\ref{eq30}) is enforced pointwise at each collocation point. This yields a system of $N_p$ nonlinear algebraic equations. Since the governing equation depends only on spatial derivatives of $h$, it is invariant under the addition of an arbitrary constant. Therefore, it admits a family of solutions which differ by their mean thickness. This degeneracy is removed by fixing the mean film thickness via the constraint 
\begin{equation}
 \int_{0}^{1} \left( h(x) - \eta f(x) \right) \, dx = 1.   
\end{equation}
The resulting nonlinear system is solved using a Newton-based root-finding algorithm (\textit{FindRoot} in \textsc{Mathematica}). The initial condition is given as a spatially nonuniform steady state obtained from the asymptotic solution derived in the previous section (equation (\ref{eq36})). Convergence is declared when the residual norm falls below $10^{-10}$. The computed solution represents a steady, spatially periodic film profile with fixed mean thickness  ($\overline{h}=1$). The steady state obtained numerically has been shown in the FIG \ref{fig:2}$(a)$ with solid lines. For small roughness amplitudes $(\eta \leq 0.1)$, the numerical solutions are in excellent agreement with the asymptotic predictions, confirming the validity of the perturbation analysis in this regime. As $\eta$ increases, quantitative deviations between the numerical and analytical solutions become apparent. This indicates the breakdown of the small-amplitude assumption in the asymptotic expansion. However, the steady-state profile remains qualitatively well captured by the asymptotic theory. \\
Surface roughness fundamentally alters the nature of the steady state. In contrast to a smooth corneal surface $(\eta=0)$, where the steady film thickness is spatially uniform, surface roughness induces a non-uniform steady state with thickness modulated at the wavelength of the roughness. The steady film profile is phase-shifted by $180 ^\circ$   relative to the imposed surface roughness as shown in FIG \ref{fig:2}. For $\eta=0$, the thickness of the tear film remains constant along the corneal surface $(h_s=1)$ and both capillary and van der Waals forces vanish simultaneously. However, for a rough corneal surface the effective film thickness $(h-\eta f(x))$ varies spatially, and the steady state is maintained by a balance between capillary and van der Waals forces.\\
FIG \ref{fig:2}$(b)$ illustrates the steady-state film thickness for different values of the reduced capillary number $(C)$. The amplitude of the film thickness decreases systematically as $C$ increases. This is because the amplitude of the fundamental mode (coefficient of $\sin(kx)$) in equation (\ref{eq36}) is $(1+B) \propto C^{-1}$. Physically, larger value of $C$ corresponds to a stronger dominance of surface tension relative to viscous effects. Surface tension resists interfacial deformation by suppressing curvature of the film. As a result, the amplitude of the film thickness decreases with increase in $C$ as shown in FIG \ref{fig:2}$(b)$. In the limit of sufficiently large $C$, the steady-state solution approaches a spatially uniform film as $(1+B) \approx 0$.

\section{Linear stability analysis}
\label{sec:LSA}

\subsection{Floquet Theory}
To examine the linear stability of the steady state, we introduce small perturbations to the film thickness, lipid concentration, and velocity components of the form
\begin{equation}
\begin{aligned}
h(t,x) &= h_s(x) + \delta h_1(t,x),\\
\gamma(t,x) &= \gamma_s + \delta \gamma_1(t,x),\\
u(t,x) &=  \delta u_1(t,x),\\
w(t,x) &=  \delta w_1(t,x).
\end{aligned}
\label{eq39}
\end{equation}

where $ \delta \ll 1$ and the quantities $h_1,\gamma_1,u_1,w_1$ denote the perturbation variables. Substitution of (\ref{eq39}) into the governing equations (\ref{eq8})-(\ref{eq17}) and retaining terms at $O(\delta)$ yield a linearized system governing the perturbations. The full set of linearized equations and boundary conditions is provided in the Supplementary Material (Section S.4). Solving linearized momentum equations allows the velocity perturbations $u_1$ and $w_1$ to be expressed in terms of the perturbations $h_1$  and $\gamma_1$. The explicit expressions for the velocity perturbations $u_1$ and $w_1$ are given in the Supplementary Material (Section S.4). Substituting these expressions into the linearized kinematic boundary condition and the lipid transport equation yields the coupled evolution equations,     \begin{subequations}\label{eq43}
\begin{align}
\frac{\partial h_1}{\partial t}
&=
P_1(x) h_1
+ Q_1(x) h_1'
+ R_1(x) h_1''
+ S_1(x) h_1'''
+ T_1(x) h_1''''
+ U_1(x) \gamma_1'
+ V_1(x) \gamma_1''
\\
\frac{\partial \gamma_1}{\partial t}
&=
P_2(x) h_1
+ Q_2(x) h_1'
+ R_2(x) h_1''
+ S_2(x) h_1'''
+ T_2(x) h_1''''
+ U_2(x) \gamma_1'
+ V_2(x) \gamma_1''
\end{align}
\end{subequations}

$ P_i (x),Q_i (x),R_i (x),T_i (x),U_i (x)$ and $V_i (x)$ for $i=1,2$ are spatially periodic coefficient functions that arise from the periodicity of the steady state. Their explicit forms are provided in Appendix \ref{App:A}.
The linearized evolution equations (\ref{eq43}) have coefficients that are periodic in the streamwise direction with the same period as that of the corneal roughness. Due to spatial periodicity, these coefficients vary in space. Hence, the classical normal-mode linear analysis is no longer applicable. Instead, we employ Floquet theory to perform linear stability of equations (\ref{eq43}). Accordingly, perturbations are sought in Bloch \citep{ajaev2013application,kuchment2012floquet} form as,
\begin{figure}
  \centering
  \includegraphics[width=1\textwidth]{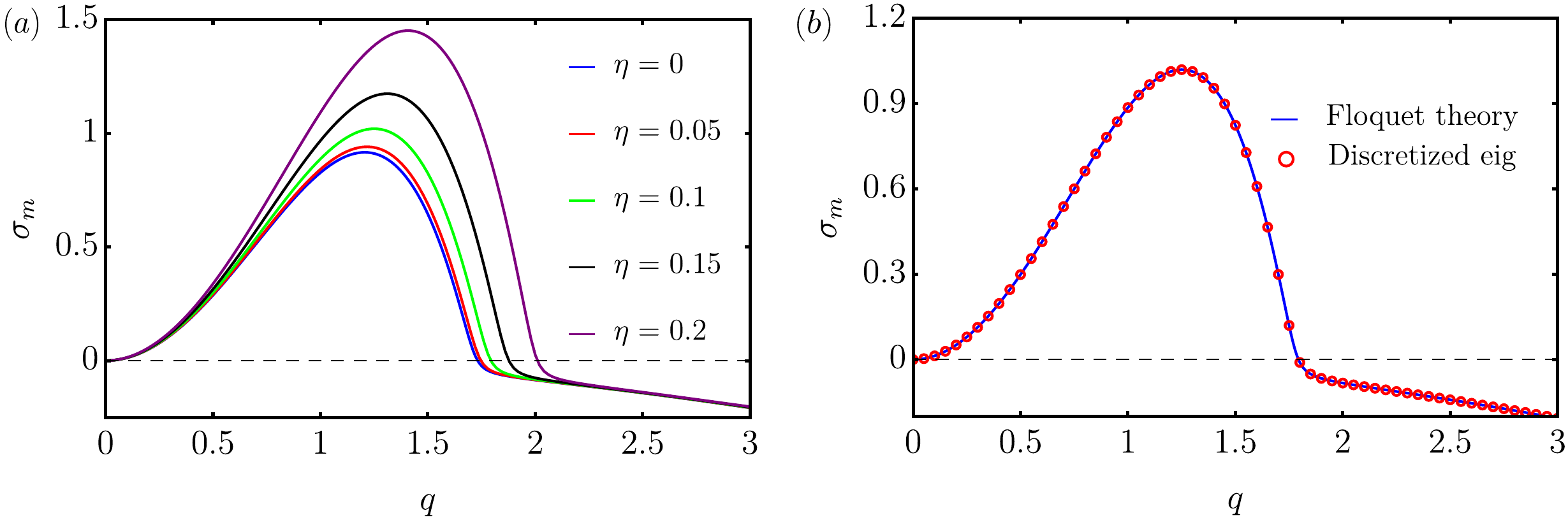}
  \caption{Variation of the perturbation growth rate $\sigma$ with wavenumber $q$. $(a)$ Dispersion curves for different values of the surface roughness amplitude $\eta$ $(b)$ Comparison between the dispersion relation obtained using Floquet theory and the discretised eigenvalue method for $\eta =0.1$. Other parameters are $C=1, M=0.1,Pe_s=100,\beta=0.1,A_k=1$ and $\gamma_s=0.5$}
  \label{fig:3}
\end{figure}
\begin{equation}
h_1 = e^{\sigma t} e^{\alpha x} \phi(x),
\label{eq45}
\end{equation}
and
\begin{equation}
\gamma_1 = e^{\sigma t} e^{\alpha x} \psi(x).
\label{eq46}
\end{equation}
where $\sigma$ denotes the temporal growth rate of the perturbation, $\alpha$ is the Floquet (Bloch) wavenumber, and $\phi(x)$ and $\psi(x)$ are complex-valued functions that are periodic with the same period as the imposed corneal roughness. The Floquet wave number must be purely imaginary for our system (Appendix \ref{App:B}) hence, $ \alpha=i q$.  The corresponding spatial derivatives are  given by
\begin{equation}
\begin{aligned}
\frac{d^{n} h_1}{dx^{n}}
&=
e^{\sigma t} e^{i q x}
\left( \frac{d}{dx} + i q \right)^{n} \phi(x),\\
\frac{d^{n} \gamma_1}{dx^{n}}
&=
e^{\sigma t} e^{i q x}
\left( \frac{d}{dx} + i q \right)^{n} \psi(x).  
\end{aligned}
\label{eq47}
\end{equation}

The exponent $n$ in equation (\ref{eq47}) denotes the order of the derivative and $\frac{d}{dx}$ represents the first order derivative. The domain $[0,1]$ is discretised using a uniform grid of $N_p$ collocation points defined by,
\begin{equation}
x_j = \frac{j}{N_p}, 
\qquad
j = 0,1,2,\ldots, N_p - 1.
\label{eq49}
\end{equation}
The coefficients $ P_i (x),Q_i (x),R_i (x),T_i (x),U_i (x)$ and $V_i (x)$ are evaluated at these collocation points. The resulting differential operators are substituted by the Fourier differentiation matrices which are constructed as described in \citep{weideman2000matlab,trefethen2000spectral}. The first-order derivative is obtained via the differentiation matrix $\mathbb{D}_1$ while higher-order derivatives are computed through matrix products,
\begin{equation}
\mathbb{D}_2 = \mathbb{D}_1\cdot \mathbb{D}_1, \mathbb{D}_3 = \mathbb{D}_2\cdot \mathbb{D}_1 \quad \text{and} \quad \mathbb{D}_4 = \mathbb{D}_2\cdot \mathbb{D}_2
\label{eq50}
\end{equation}

Substituting the Bloch ansatz into equations (\ref{eq43}) yields an eigenvalue problem for the growth rate $\sigma$ parameterized by the Bloch wavenumber $q$. The resulting generalized eigenvalue problem can be written as,
\begin{equation}
\sigma \mathbf{Z} =
\begin{pmatrix}
A_{h\phi}(q) & A_{h\psi}(q) \\
A_{\gamma\phi}(q) & A_{\gamma\psi}(q)
\end{pmatrix}
\mathbf{Z}
\label{eq51}
\end{equation}
Here, $\mathbf{Z}=\left[\phi_1,\, \phi_2,\, \ldots,\, \phi_{N_p}, \psi_1,\, \psi_2,\, \ldots,\, \psi_{N_p}\right]^{T}$ is the vector of unknown variables evaluated at the collocation points. The matrices $A_{h\phi}(q)$ and $A_{h\psi}(q)$ (each of size $N_p \times N_p$) arise from the coefficients of $\phi$ and $\psi$ in equation (\ref{eq43}$a$). Similarly, the matrices $A_{\gamma\phi}(q)$ and $A_{\gamma\psi}(q)$ are obtained from the coefficients of $\phi$ and $\psi$ in equation (\ref{eq43}$b$). For each prescribed value of $q$, the resulting eigenvalue problem (equation (\ref{eq51})) admits a discrete spectrum of eigenvalues. The dominant eigenvalue, $\sigma_m = \max(Re(\sigma))$ determines the stability of the corresponding wave number. Here, $Re(\sigma)$ denotes the real part of the complex eigenvalue $\sigma$. The dispersion relation $\sigma_m (q)$ is obtained by repeating procedure over a range of $q$ and identifying the leading growth rate for each $q$. \\

FIG \ref{fig:3}$(a)$ shows the dependence of the dominant growth rate $\sigma_m$ on the wavenumber $q$ for different values of the surface roughness amplitude $\eta$. As $\eta$ increases, the maximum growth rate increases. This indicates that surface roughness enhances the instability of the tear film. We note that that in the limit $\eta=0$, the dispersion relation obtained from Floquet theory reduces exactly to that predicted by classical normal-mode linear stability analysis. The most unstable wavenumber $q_m$ is defined as the value of $q$ at which the growth rate attains its maximum i.e.  $\left.\frac{d\sigma_m}{dq}\right|_{q_m} = 0$. The cutoff wave number $(q_c )$ is defined as the wave number across which the growth rate changes its sign. Both $q_m$ and $q_c$ depend on the roughness amplitude $(\eta)$ and increase as $\eta$ increases. The increase in $q_m$ indicates that substrate roughness not only amplifies the instability but also shifts the characteristic instability length scale $\frac{2\pi}{q_m}$ . Next, we verify our linear stability results using the discretized eigenvalue approach.

\subsection{Discretized eigenvalue approach}
To validate the dispersion curves from Floquet theory, we analyse the coupled linear system (\ref{eq43}) using  an alternative  approach. This is based on a Fourier-Galerkin expansion to compute discrete eigenvalues. Since the substrate pattern is periodic, we consider an extended computational domain of length $(L_e >> l)$ consisting of $N_g$ repeated unit cells such that $L_e=l \times N_g$. Here $l=1$ is the fundamental wavelength of substrate pattern. The perturbations $h_1$ and $\gamma_1$ are expanded as Fourier series over the extended domain $L_e$, 

\begin{equation}
\begin{aligned}
h_1
&=
\sum_{n=1}^{\infty}
\left(
h_n e^{i q_n x}
+
\bar{h}_n e^{-i q_n x}
\right),\\
\gamma_1
&=
\sum_{n=1}^{\infty}
\left(
\gamma_n e^{i q_n x}
+
\bar{\gamma}_n e^{-i q_n x}
\right).
\label{eq52d}
\end{aligned}
\end{equation}

Here, $q_n=\frac{2 \pi n}{L_e}$   are the discrete wavenumbers. $\bar{h}_n$, $\bar{\gamma}_n$ denotes the complex conjugate of $h_n$ and $\gamma_n$ respectively. 
The spatially periodic coefficient $P_i (x)$ for $i=1,2$ are similarly expanded as,

\begin{equation}
P_i(x)
=
\sum_{j=0}^{\infty}
\left(
P_{ij} e^{i q_j x}
+
\text{c.c.}
\right)
\label{eq54d}
\end{equation}

where $q_j=\frac{2 \pi j}{L_e}N_g=\frac{2 \pi j}{l}$ and the Fourier coefficients $(P_{ij})$ are computed by,
\begin{equation}
P_{ij}
=
 \int_{0}^{1}
P_i(x)\, e^{-i q_j x}\, dx
\label{eq55d}
\end{equation}
All remaining periodic coefficient $Q_i (x),R_i (x),T_i (x),U_i (x)$ and $V_i (x)$ are expanded analogously as equation (\ref{eq54d}). These coefficients are evaluated at the steady-state solution $h_s (x)$ which is a smooth function of $x$. Hence, the Fourier coefficients decay rapidly with $j$. This ensures fast convergence of the truncated expansions. We substitute the Fourier expansions (\ref{eq52d}) into the linearized equations (\ref{eq43}). Multiplying by $e^{-iq_m x}$, and integrating over the full computational domain $[0,L_e]$, we exploit the orthogonality relation
\begin{equation}
    \frac{1}{L_e}\int_{0}^{L_e}
 e^{-i( q_a - q_b )x}\, dx =\delta_{ab}
\label{eq56d}
\end{equation}
to project onto each Fourier mode $m$. After projection, the non-zero contributions to row $m$ arise only from those modes $n$ which satisfy one of the following resonance conditions:
\begin{equation}
\begin{aligned}
n &= m, && \text{self-coupling},\\
n &= m \pm N_g j, && \text{forward and backward coupling},\\
n &= N_g j - m, && \text{cross coupling to conjugate modes}.
\end{aligned}
\label{eq:modecoupling}
\end{equation}
$j_\text{Max}$ is chosen sufficiently large to ensure convergence. In the present computations, we use $j_\text{Max} \ge 200$. After truncation to $N_p$ modes, the system reduces to a finite-dimensional eigenvalue problem. Since the coefficients are generally complex, the Fourier amplitudes $h_n,\gamma_n$ and their complex conjugates  $\bar{h}_n, \, \bar{\gamma}_n$ must be treated as independent variables. We therefore introduce a vector,
\begin{equation}
\mathbf{Z} =
\left[
h_1, h_2, h_3, \ldots, h_{N_p},
\bar{h}_1, \bar{h}_2, \bar{h}_3, \ldots, \bar{h}_{N_p},
\gamma_1, \gamma_2, \gamma_3, \ldots, \gamma_{N_p},
\bar{\gamma}_1, \bar{\gamma}_2, \bar{\gamma}_3, \ldots, \bar{\gamma}_{N_p}
\right]^T
\label{eq60d}
\end{equation}
The projected linearized equations (\ref{eq43}) for the film thickness and surfactant concentration amplitudes take the form:
\begin{equation}
\begin{aligned}
\sum_{n=1}^{N_p} \frac{d h_n}{dt}
&=
\sum_{n=1}^{N_p} A_{hh}(m,n)\, h_n
+
\sum_{n=1}^{N_p} A_{h\bar{h}}(m,n)\, \bar{h}_n
+
\sum_{n=1}^{N_p} A_{h\gamma}(m,n)\, \gamma_n
+
\sum_{n=1}^{N_p} A_{h\bar{\gamma}}(m,n)\, \bar{\gamma}_n,\\
\sum_{n=1}^{N_p} \frac{d \gamma_n}{dt}
&=
\sum_{n=1}^{N_p} A_{\gamma h}(m,n)\, h_n
+
\sum_{n=1}^{N_p} A_{\gamma \bar{h}}(m,n)\, \bar{h}_n
+
\sum_{n=1}^{N_p} A_{\gamma \gamma}(m,n)\, \gamma_n
+
\sum_{n=1}^{N_p} A_{\gamma \bar{\gamma}}(m,n)\, \bar{\gamma}_n,    
\end{aligned}
\label{eq61d}
\end{equation}
for $m=1,2,3,…..N_p$. Taking the complex conjugate of equations (\ref{eq61d}) yields two additional evolution equations for $\bar{h}_n$ and  $\bar{\gamma}_n$. The full linearized system may then be written compactly as,
\begin{equation}
    \frac{d \mathbf{Z}}{dt}= \mathbf{A} \, \mathbf{Z}
    \label{eq62d}
\end{equation}
where $\mathbf{A}$ is a $4N_p \times 4N_p$ block matrix composed of the coupling matrices $A_{hh}, A_{h\bar{h}}, A_{h\gamma}, A_{h\bar{\gamma}} $ and $A_{\gamma h}, A_{\gamma \bar{h}}, A_{\gamma \gamma}, A_{\gamma \bar{\gamma}} $ are given by,
\begin{equation}
\mathbf{A} =
\begin{pmatrix}
A_{hh} & A_{h\bar{h}} & A_{h\gamma} & A_{h\bar{\gamma}} \\
\overline{A_{h\bar{h}}} & \overline{A_{hh}} & \overline{A_{h\bar{\gamma}}} & \overline{A_{h\gamma}} \\
A_{\gamma h} & A_{\gamma \bar{h}} & A_{\gamma \gamma} & A_{\gamma \bar{\gamma}} \\
\overline{A_{\gamma \bar{h}}} & \overline{A_{\gamma h}} & \overline{A_{\gamma \bar{\gamma}}} & \overline{A_{\gamma \gamma}}
\end{pmatrix}
\label{eq63d}
\end{equation}

The diagonal blocks corresponding to the self-coupling condition $n=m$ are

\begin{equation}
\begin{aligned}
A_{hh}(m,m) &= P_{10}+iQ_{10}q_n-R_{10}q_n^2-iS_{10}q_n^3+T_{10}q_n^4,\\
A_{h\gamma}(m,m) &= iU_{10}q_n-V_{10}q_n^2,\\
A_{\gamma h}(m,m) &= P_{20}+iQ_{20}q_n-R_{20}q_n^2-iS_{20}q_n^3+T_{20}q_n^4,\\
A_{\gamma\gamma}(m,m) &= iU_{20}q_n-V_{20}q_n^2.
\end{aligned}
\end{equation}

Off-diagonal blocks arise from mode coupling through shifted harmonics $k_1$, $k_2$, and $k_3$.

For $n=k_1=m-jN_g$, if $1\le k_1\le N_p$,
\begin{equation}
\begin{aligned}
A_{hh}(m,k_1) &= P_{1j}+iQ_{1j}q_n-R_{1j}q_n^2-iS_{1j}q_n^3+T_{1j}q_n^4,\\
A_{h\gamma}(m,k_1) &= iU_{1j}q_n-V_{1j}q_n^2,\\
A_{\gamma h}(m,k_1) &= P_{2j}+iQ_{2j}q_n-R_{2j}q_n^2-iS_{2j}q_n^3+T_{2j}q_n^4,\\
A_{\gamma\gamma}(m,k_1) &= iU_{2j}q_n-V_{2j}q_n^2.
\end{aligned}
\label{eq67}
\end{equation}

For $n=k_2=m+jN_g$, if $1\le k_2\le N_p$,
\begin{equation}
\begin{aligned}
A_{hh}(m,k_2) &= \overline{P}_{2j}+i\overline{Q}_{2j}q_n-\overline{R}_{2j}q_n^2-i\overline{S}_{2j}q_n^3+\overline{T}_{2j}q_n^4,\\
A_{h\gamma}(m,k_2) &= i\overline{U}_{1j}q_n-\overline{V}_{1j}q_n^2,\\
A_{\gamma h}(m,k_2) &= \overline{P}_{2j}+i\overline{Q}_{2j}q_n-\overline{R}_{2j}q_n^2-i\overline{S}_{2j}q_n^3+\overline{T}_{2j}q_n^4,\\
A_{\gamma\gamma}(m,k_2) &= i\overline{U}_{2j}q_n-\overline{V}_{2j}q_n^2.
\end{aligned}
\label{eq69}
\end{equation}

For $n=k_3=jN_g-m$, if $1\le k_3\le N_p$,
\begin{equation}
\begin{aligned}
A_{h\bar h}(m,k_3) &= P_{1j}-iQ_{1j}q_n-R_{1j}q_n^2+iS_{1j}q_n^3+T_{1j}q_n^4,\\
A_{h\bar\gamma}(m,k_3) &= -iU_{1j}q_n-V_{1j}q_n^2,\\
A_{\gamma\bar h}(m,k_3) &= P_{2j}-iQ_{2j}q_n-R_{2j}q_n^2+iS_{2j}q_n^3+T_{2j}q_n^4,\\
A_{\gamma\bar\gamma}(m,k_3) &= -iU_{2j}q_n-V_{2j}q_n^2,
\end{aligned}
\label{eq75}
\end{equation}
We note that the sign change in the odd-derivative terms in equations (\ref{eq75}) relative to equations (\ref{eq67} and \ref{eq69}) reflects the fact that the conjugate modes $\bar{h}_n$ and $\bar{\gamma}_n$ carry wavenumber $-q_n$. Consequently, the $n$\text{th} derivative contributes a factor $(-iq_n)^n$ rather than $(iq_n)^n$.

Finally, the stability problem is obtained by assuming normal modes of the form
\begin{equation}
\mathbf{Z} = \mathbf{Z_0} e^{\sigma t},
\end{equation}
which yields the eigenvalue problem
\begin{equation}
\sigma \mathbf{Z_0} = \mathbf{A} \mathbf{Z_0} .
\end{equation}

Here, the real part of $\sigma$ represents the temporal growth rate. The eigenvalues and corresponding eigenvectors of the matrix $A$ are obtained numerically using the \textit{Eigensystem} command in \textsc{Mathematica}.

To construct the dispersion relation, each eigenmode is assigned a dominant wavenumber by identifying the Fourier component with the largest amplitude within the corresponding eigenvector. Specifically, for the $j$th eigenvector $\mathbf{Z_0}^{(j)}$ of dimension $4N_p$, we compute
\begin{equation}
n^* = \text{arg}\,\max\limits_{1 \le n \le 4N_p} \left| \mathbf{Z_0}^{(j,n)} \right|,
\end{equation}
where $\mathbf{Z_0}^{(j,n)}$ denotes the $n$th component of the $j$th eigenvector.

Since the state vector $\mathbf{Z_0}$ consists of four blocks of size $N_p$, the dominant index $n^*$ is mapped to the corresponding physical wavenumber as
\begin{equation}
q =
\begin{cases}
q_{n^*} &  \text{if} \quad 1 \le n^* \le N_p, \\
q_{n^* - N_p}  & \text{if} \quad N_p < n^* \le 2N_p, \\
q_{n^* - 2N_p} & \text{if} \quad 2N_p < n^* \le 3N_p, \\
q_{n^* - 3N_p} & \text{if} \quad 3N_p < n^* \le 4N_p .
\end{cases}
\end{equation}

The dispersion curve is then constructed by plotting the maximum growth rate
\[
\sigma_m = \max \bigl(\mathrm{Re}(\sigma)\bigr) \ge -0.25
\]
against its dominant wavenumber $q$. Convergence of the results with respect to the truncation parameters $N_p$ and $j_{\max}$ is verified by systematically increasing their values until the change in $\sigma_m$ is less than $10^{-3}$.

FIG \ref{fig:3}$(b)$ compares the dispersion relation obtained using Floquet theory with that computed via the discretised eigenvalue method for $\eta=0.1$. The two approaches are in excellent agreement across the range of wavenumbers considered. This validates the linear stability analysis obtained from the two independent formulations. The discretized eigenvalue framework generalizes classical normal-mode stability analysis by explicitly accounting for the spatially periodic coefficients. It naturally captures mode-coupling effects that are absent in spatially homogeneous systems.

\begin{figure}
  \centering
  \includegraphics[width=1\textwidth]{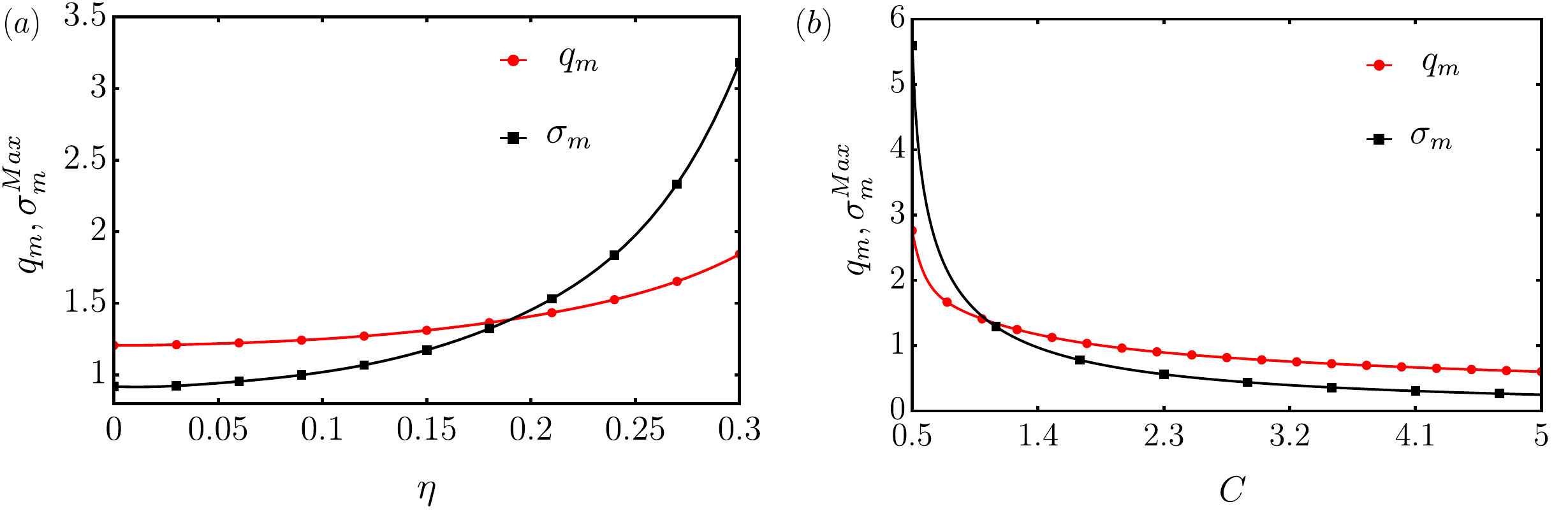}
  \caption{Variation of the maximum growth rate $\sigma_m^{Max}$ and the most unstable wavenumber with $(a)$ surface roughness amplitude $(\eta)$ and $(b)$ reduced capillary number $C$ for $\eta=0.2$. Other parameters are $C=1, M=0.1,Pe_s=100,\beta=0.1,A_k=1$ and $\gamma_s=0.5$.}
  \label{fig:4}
\end{figure}

\begin{figure}
  \centering
  \includegraphics[width=1\textwidth]{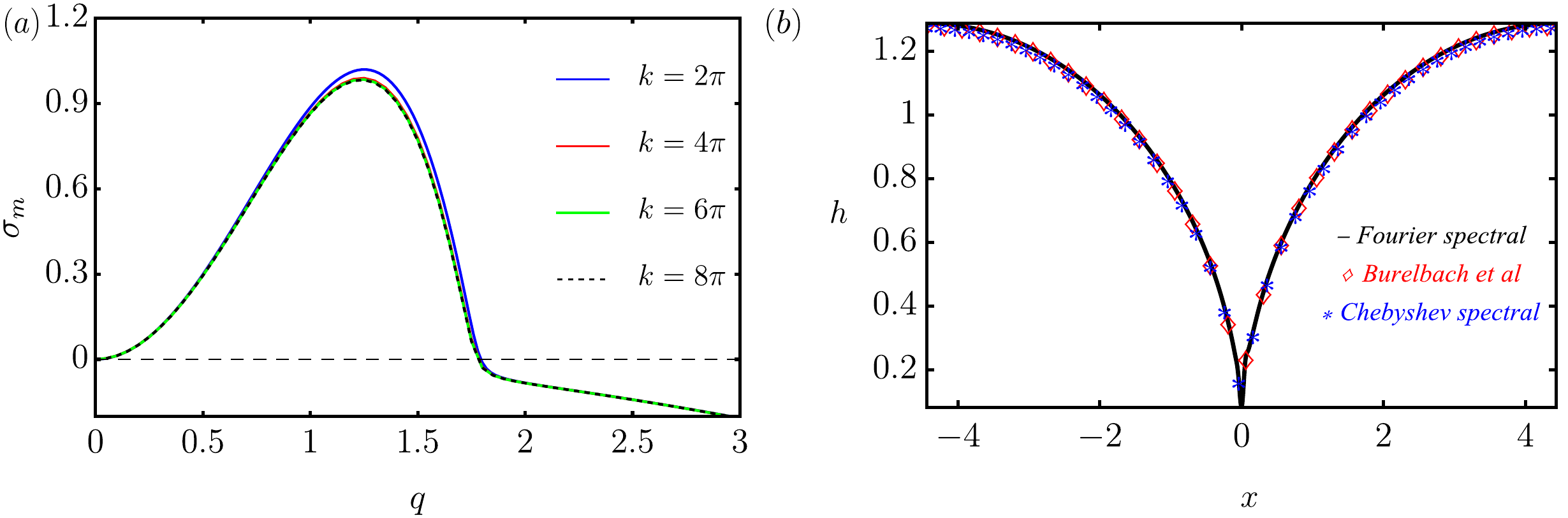}
  \caption{$(a)$ Effect of wavenumber of the roughness on the linear stability characteristics $(b)$ Validation of our in-house numerical scheme using the Fourier spectral method and the Chebyshev spectral method implemented in \textsc{Mathematica}. The figure shows the film profile at the time of rupture, demonstrating excellent agreement between the two methods and the result of Burelbach \textit{et al} \cite{burelbach1988nonlinear} }
  \label{fig:5}
\end{figure}

FIG \ref{fig:4}$(a,b)$ illustrate the dependence of the maximum growth rate $\sigma_m^{Max} = \max(\sigma_m)$
 and the most unstable wavenumber $q_m$ on the surface roughness amplitude $\eta$ and reduced Capillary number $C$. Both $\sigma_m$ and $q_m$ increase monotonically with $\eta$. The increase in $\sigma_m$ indicates that surface roughness enhances the instability. Roughness induces spatial modulation of the base-state film thickness, which locally strengthens destabilising van der Waals forces and increases the sensitivity of the film to perturbations. Consequently, disturbances amplify more rapidly as $\eta$ increases. The increase in $q_m$ indicates a shift of the instability towards shorter wavelengths as shown in FIG \ref{fig:4}$(a)$.\\
 In contrast, both $\sigma_m$ and $q_m$ decrease with increasing $C$. Since $C$ measures the relative influence of surface tension and viscous forces, larger values correspond to stronger capillary stabilization. Surface tension suppresses interfacial curvature and damps short-wavelength disturbances. Hence, the overall growth rate decreases. As a result, the instability weakens and shifts towards longer wavelengths as $C$ increases as shown in FIG \ref{fig:4}$(b)$. FIG \ref{fig:5}(a) shows the dispersion curves for four different values of the roughness wavenumber. It is observed that the maximum growth rate and the corresponding most unstable wavenumber do not change significantly as the roughness frequency varies. This indicates that the stability characteristics depend only weakly on the frequency of the wall roughness. We next perform nonlinear numerical simulations to examine the spatio-temporal evolution of the tear film over the rough corneal surface.

\section{Nonlinear numerical simulation}
\label{sec:nonlin}
Two distinct wavenumbers arise in the present problem: the substrate wavenumber $k$, associated with the corneal surface roughness and the most unstable wavenumber $q_m$, obtained from the linear stability analysis. For nonlinear simulations, the computational domain length must be chosen such that it can accommodate the dominant unstable mode predicted by the linear stability analysis. In particular, if the most unstable wavenumber is $q_m$, then the corresponding wavelength is $\lambda_m=\frac{2\pi}{q_m}$. The computational domain length $\Omega$ must satisfy $\Omega \geq \lambda_m$  to allow at least one full period of this fastest-growing disturbance to exist. If $\Omega < \lambda_m$, the numerical domain artificially suppresses or distorts the growth of the physically relevant instability, leading to incorrect nonlinear evolution. Nonlinear numerical simulation to determine $h$ and $\gamma$ are performed using the Fourier spectral method over the computational domain $[0,\Omega]$. The domain length $\Omega$ is chosen as the smallest integer exceeding $\lambda_m$ to ensure that the computational domain accommodates at least one full wavelength of the most unstable mode.\\
\begin{figure}
  \centering
  \includegraphics[width=1\textwidth]{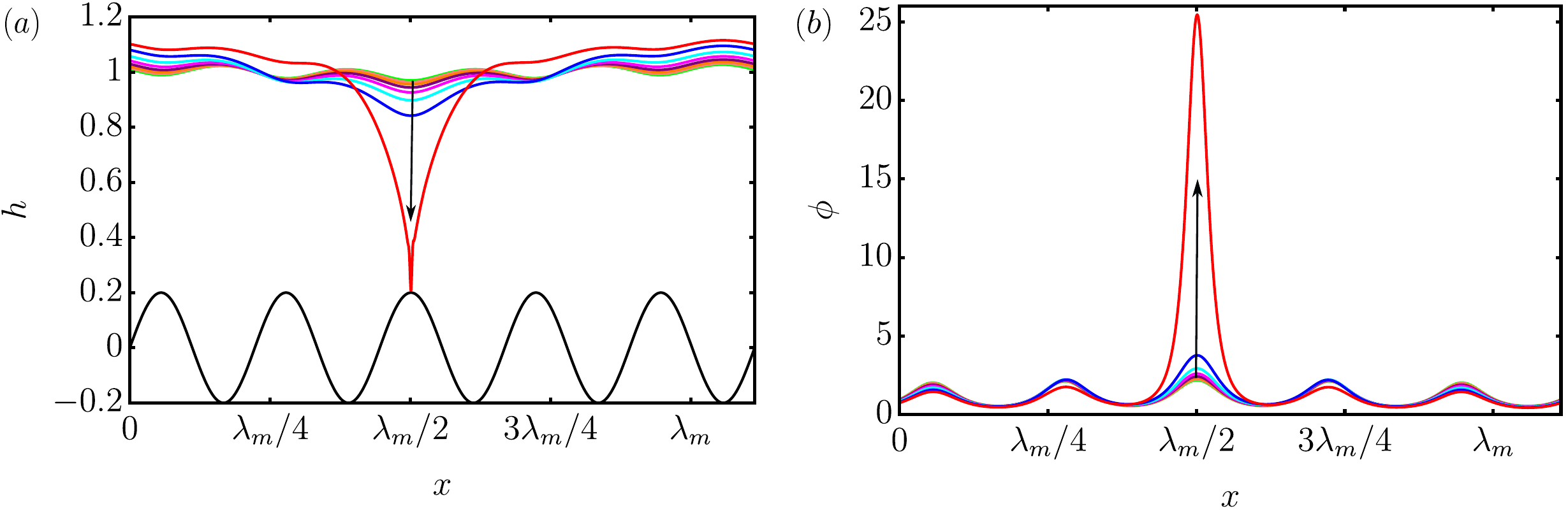}
  \caption{Temporal evolution of $(a)$ the tear-film profile obtained from the nonlinear simulations of equations (\ref{eq28}-\ref{eq29}) $(b)$ the corresponding van der Waals potential over a rough corneal surface with $\eta=0.2$. The initial condition is $h=h_s(x) +H_0 \cos (q_m x)$. The arrows indicate the progression of time from $t=0$ to the rupture time $t_\text{rup}$. The other parameters are $C=1,M=0.1,Pe_s=100,\beta=0.1,A_k=1$ and $\gamma_s=0.5.$}
  \label{fig:6}
\end{figure}
The coupled governing partial differential equations are solved numerically using a Fourier spectral collocation method on a periodic domain $[0,\Omega]$. The spatial grid consists of $N_p$ uniformly distributed points
\begin{equation}
x_j = \frac{j \Omega}{N_p}, 
\qquad
j = 0,1,2,\ldots, N_p - 1.
\label{eq52}
\end{equation}
Spatial derivatives are evaluated using Fourier differentiation matrices as described in section \ref{sec:LSA}. The spatial discretisation reduces equations (\ref{eq28}-\ref{eq29}) to a system of ordinary differential equations in time, which is advanced using the adaptive time-stepping solver \textit{NDSolve} in \textsc{Mathematica}.\\
To benchmark our code, we considered the isothermal thin film rupture problem without surfactant analysed in  \cite{burelbach1988nonlinear}. Using the Fourier spectral method, the computed rupture time is $t_\text{rup}=4.085$. For validation, we independently repeated the calculation using a Chebyshev spectral collocation method, obtaining $ t_\text{rup}=4.078$. Both values are in close agreement with the rupture time $t_\text{rup}=4.164$ reported by Burelbach \textit{et al} \cite{burelbach1988nonlinear}. This validates the accuracy and reliability of our numerical scheme. The film profiles computed using both the methods and the literature at the time of rupture are shown in FIG \ref{fig:5}$(b)$ and exhibit excellent agreement. All subsequent simulations are carried out using the Fourier spectral method due to its computational efficiency and high spatial accuracy for periodic domains. \\
The evolution equations for $h$ and $\gamma$ describing our system are integrated in time with periodic boundary conditions
\begin{equation}
\begin{aligned}
&h(t,0) = h(t,\Omega), 
\qquad
h'(t,0) = h'(t,\Omega), \\
&h''(t,0) = h''(t,\Omega), 
\quad
h'''(t,0) = h'''(t,\Omega), \\
&\gamma(t,0) = \gamma(t,\Omega), 
\qquad
\gamma'(t,0) = \gamma'(t,\Omega).
\end{aligned}
\label{eq53}
\end{equation}
Here, the prime $(')$ denotes differentiation with respect to $x$. The initial conditions are specified as $h(t=0) =h_s +H_0 \cos (q_m x)$ where $h_s$ is the steady state and $\gamma(t=0) =\gamma_s$. $q_m$  is the wavenumber corresponding to the most unstable mode identified from the linear stability analysis. For these simulations, the following baseline parameter values are used: $A_k=1,M=1,Pe_s=100,\beta=0.1,C=1,\gamma_{s}=0.5$,  and $H_0=0.01$. FIG \ref{fig:6}$(a)$ illustrates the nonlinear evolution of the tear film over a rough corneal surface for $\eta=0.2$. For these parameters, the most unstable wave number is $q_m =1.4$ and the corresponding growth rate of perturbation $\sigma_m=1.45$ is positive indicating instability. FIG \ref{fig:6}$(a)$ shows the temporal evolution of the film thickness $h$ at different time instants with a time interval of $\delta t=0.25$. The solid black line denotes the rough corneal surface. Tear film rupture is identified when the film first contacts this surface. The perturbation in the  initial condition creates a pressure gradient inside the film. This is induced by the attractive van der Waals forces which increase in magnitude as the film thins. Since these forces scale inversely with the cube of the local film thickness ($ \sim h^{-3}$), their strength grows rapidly as $h$ decreases over time as shown in FIG \ref{fig:6}$(b)$. Consequently, the rate of thinning accelerates with time. The tear film is locally thinner above the elevated regions of the corneal surface at $x=\lambda_m/2$. As a result, the van der Waals attraction is stronger in this region. The enhanced attraction increases the magnitude of the van der Waals forces, which pull the film downward more strongly. Hence, localized thinning accelerates the rupture process leading to film rupture at $x= \lambda_m/2$ and $t_\text{rup}=1.71$.\\

\begin{figure}
  \centering
  \includegraphics[width=1\textwidth]{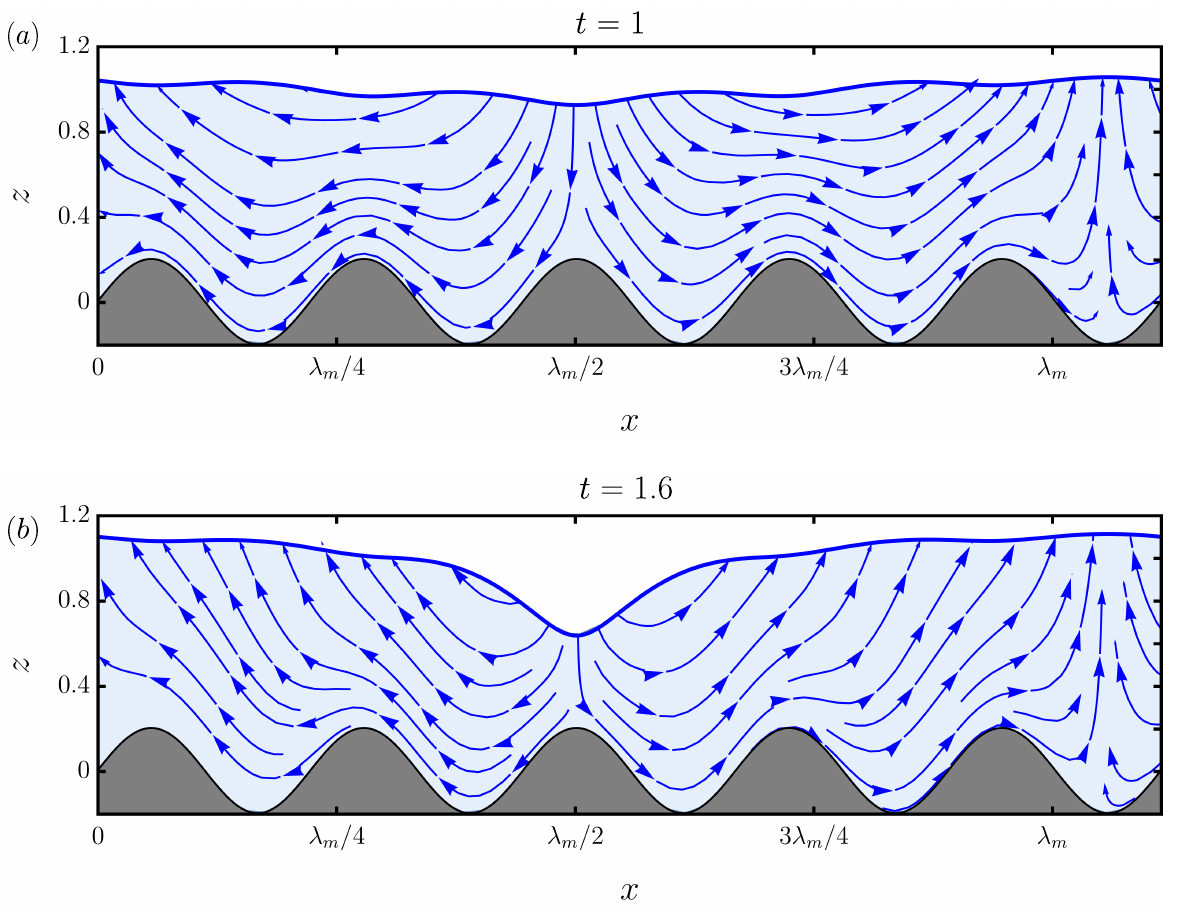}
  \caption{Streamlines illustrating the flow field within the tear film over a rough corneal surface $\eta=0.2$ at $(a)$ $t=1$ and $(b)$ $t=1.6$. The other parameters are $C=1,M=0.1,Pe_s=100,\beta=0.1,A_k=1$ and $\gamma_s=0.5.$}
  \label{fig:7}
\end{figure}
The van der Waals forces drive fluid from thinner regions (valleys) toward thicker regions (crests), causing the valleys to thin further over time. FIG \ref{fig:7}$(a,b)$ illustrates the streamlines of the fluid motion at time $ t=1$ and $t=1.6$ respectively in the original coordinate $(x,z)$. At $t=1$, the fluid motion is directed away from the locally thinned region located at $x=\lambda_m/2$ and toward the relatively thicker regions on either side. The fluid velocity is relatively small at this time. The viscous resistance causes the flow to closely follow the rough corneal surface, as shown in FIG \ref{fig:7}$(a)$. Hence, thinning at this stage remains smooth and the streamlines indicate a gradual outward drainage from the thinning line $x=\lambda_m/2$. At the tear-air interface, the velocity vectors on both sides of the thinning zone point vertically upward. We emphasize that this does not indicate mass loss from the interface, rather it reflects the local upward motion of the interface on both sides. This is confirmed in \ref{fig:7}$(b)$, where the interface is visibly elevated in these regions. The upward displacement arises from volume conservation. As the film at $x=\lambda_m/2$ depresses under the action of van der Waals forces, the fluid is redistributed laterally. This causes the adjacent regions to rise. At a later time $t=1.6$, the flow pattern remains qualitatively similar; however, the streamlines become more elongated in the vertical direction. The upward motion of the fluid is stronger and the flow follows the wavy wall to a lesser depth compared to $t=1$. The streamlines for smooth corneal surface has been shown in the Supplementary Material (Figure S1).\\
\begin{figure}
  \centering
  \includegraphics[width=1\textwidth]{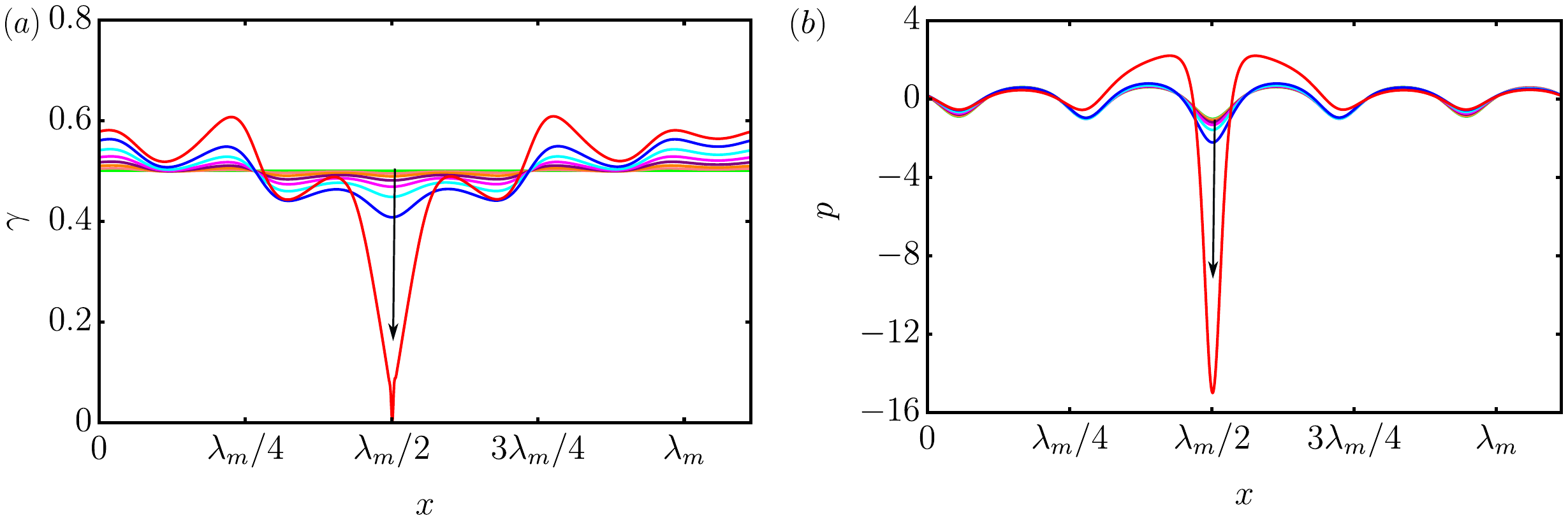}
  \caption{ Temporal evolution of $(a)$ the lipid concentration and $(b)$ the pressure distribution over a rough corneal surface with  $\eta=0.2$. The initial condition is $h=h_s(x) +H_0 \cos (q_m x)$ and $\gamma=\gamma_s$. The arrows indicate the progression of time from $t=0$ to the rupture time $t_\text{rup}$. The other parameters are $C=1,M=0.1,Pe_s=100,\beta=0.1,A_k=1$ and $\gamma_s=0.5.$}
  \label{fig:8}
\end{figure}

The accompanying fluid motion also transports lipid molecules from the valleys to the crests. This advection leads to a reduction in lipid concentration at the valleys and an accumulation at the crests as shown in FIG \ref{fig:8}$(a)$. The resulting nonuniform surfactant distribution creates surface tension gradients causing higher surface tension in the thinner regions and lower surface tension in the thicker regions. These gradients give rise to Marangoni stresses that drive fluid from the crests toward the valleys, opposing the flow induced by van der Waals forces. Consequently, the lipid redistribution acts as a stabilizing mechanism, which resist further thinning of the tear film. Additionally, capillary forces arising from the curvature gradients in the film act to resist rupture. However, the stabilizing Marangoni and capillary effects are not strong enough to counterbalance the dominant van der Waals attraction. As a result, despite the local stabilization imparted by the lipid gradients, the film continues to thin and eventually ruptures when the minimum film thickness reaches the corneal surface. Following rupture, exposure of the lipid layer creates a locally hydrophobic surface. This suppresses rewetting and stabilises the dry patch formation over the corneal surface. FIG \ref{fig:8}$(b)$ shows the evolution of the pressure distribution over the rough corneal surface from $t=0$ to $t=t_\text{rup}$ with time interval $ \delta t=0.25$. The pressure within the tear film is given by $p=-C \frac{\partial^2 h}{\partial x^2}$. In the vicinity of $x=\lambda_m/2$, the film thickness attains a local minimum at each time instant. Consequently, $\frac{\partial^2 h}{\partial x^2}>0$ in this region. As thinning progresses, the curvature $\frac{\partial^2 h}{\partial x^2}$ increases in magnitude leading to a progressively more negative pressure, as shown in FIG \ref{fig:8}$(b)$. The resulting pressure gradients drive fluid away from the thinning region and strengthens the outward flux of tear fluid from this region. This leads to local depletion of tear fluid in the thinning region. Simultaneously, the van der Waals attraction strengthens due to its $h^{-3}$ dependence. This further intensify the local negative pressure. The combined amplification of curvature and intermolecular attraction establishes a nonlinear positive feedback mechanism. The thinning increases the destabilising pressure, which in turn accelerates further thinning.\\

\begin{figure}
  \centering
  \includegraphics[width=1\textwidth]{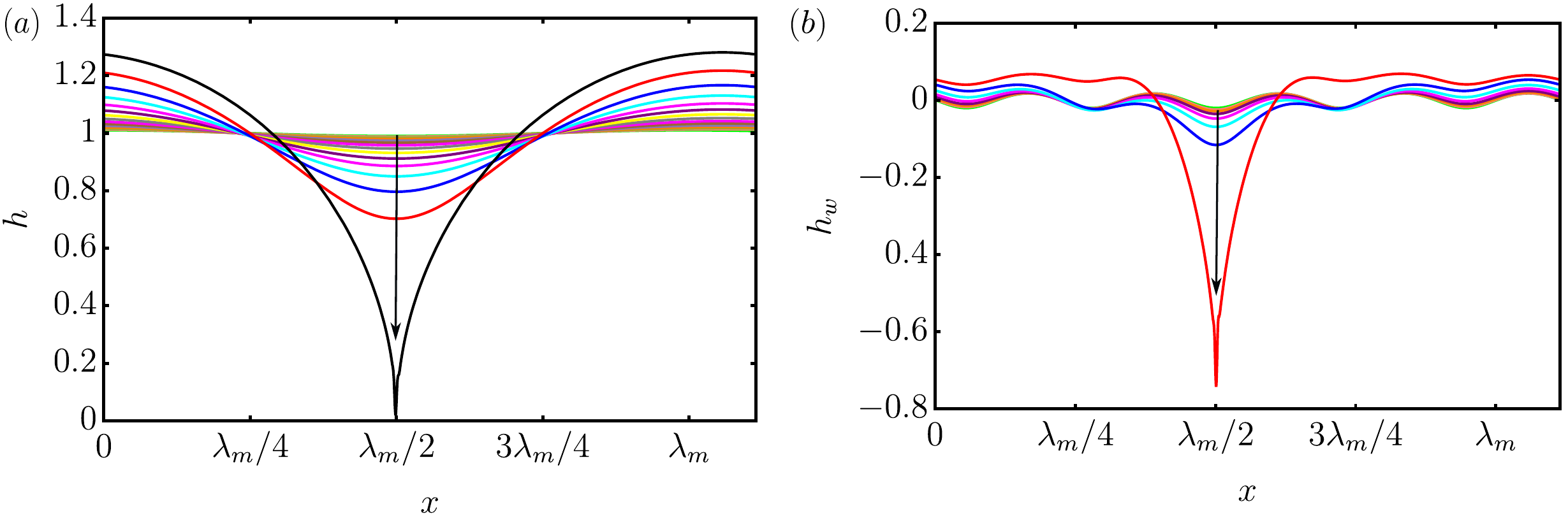}
  \caption{ Temporal evolution of $(a)$ the film thickness over a smooth corneal surface $(\eta=0)$ and $(b)$ the effect of substrate roughness $(\eta=0.2)$ on tear-film evolution. The initial condition is  $h=h_s(x) +H_0 \cos (q_m x)$ and $\gamma=\gamma_0$. The arrows indicate the progression of time from $t=0$ to the rupture time $t_\text{rup}$. The other parameters are $C=1,M=0.1,Pe_s=100,\beta=0.1,A_k=1$ and $\gamma_s=0.5.$}
  \label{fig:9}
\end{figure}

The evolution of the film over the smooth corneal surface $(\eta=0)$ is shown in FIG \ref{fig:9}$(a)$. In this case, steady state thickness is $h_s=1$ and the most unstable wave number $q_m=1.21$. The initial condition is therefore perturbed with the wavelength associated with this mode. For the smooth cornea, rupture is considered to occur when the local film thickness reaches $z=0$. The tear film ruptures at $t=3.59$ which is significantly higher than that of the rough corneal surface where rupture occurs at $t=1.71$ as shown in FIG \ref{fig:6}. This confirms that surface roughness amplifies the destabilising mechanisms and promotes earlier tear-film breakup. The arrow in the figure indicates the progression of time from $t=0$ to $t_\text{rup}$ with a time increment of $\delta t=0.25$. \\

To understand the effect of corneal roughness, we define a variable $h_w=h|_\eta-h|_{\eta=0}$ which represents the deviation in film thickness induced solely by the corneal roughness. Here, $h|_{\eta}$ represents the film thickness at a given time for non-zero roughness amplitude $\eta$ while $h|_{\eta=0}$ corresponds to the evolution over a smooth corneal surface. FIG \ref{fig:9}$(b)$ illustrates the spatiotemporal evolution of $h_w$. As time progresses, $h_w$ decreases and attains its minimum  above the elevated surface at the onset of rupture. This indicates that surface roughness increasingly promotes local thinning of the film above the elevated portions of the wall and accelerates the rupture process.\\

\begin{figure}
  \centering
  \includegraphics[width=1\textwidth]{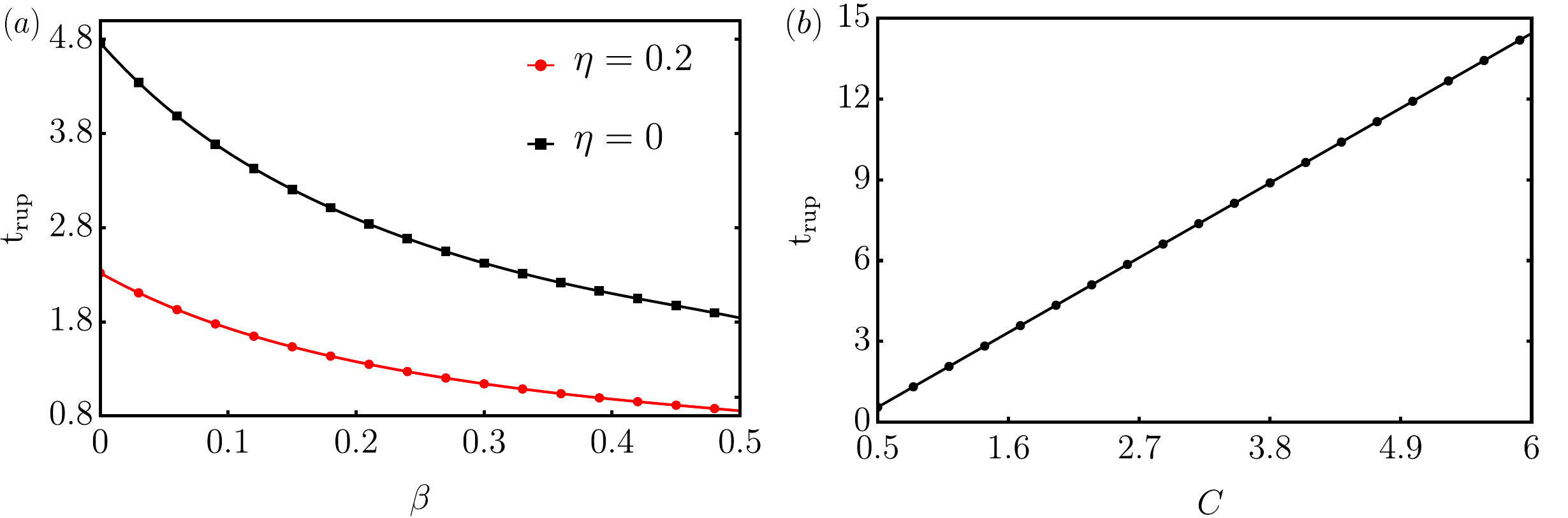}
  \caption{ Dependence of tear breakup time on $(a)$ slip coefficient $\beta$ $(b)$ the reduced Capillary number $C $ for $\eta=0.2$. The other parameters are $C=1,M=0.1,Pe_s=100,\beta=0.1,A_k=1$ and $\gamma_s=0.5.$ }
  \label{fig:10}
\end{figure}

Recent evidences suggest that the ocular surface responds dynamically to both physiological and pathological conditions, which can significantly alter the slip coefficient and surface roughness of the cornea \citep{king2014tear,liu1999corneal}. FIG \ref{fig:10}$(a)$ illustrates the influence of the slip coefficient $(\beta)$ on the tear breakup time $(t_\text{rup})$ for both smooth and rough corneal surfaces. In both cases, $t_\text{rup}$ decreases monotonically with increasing $\beta$. This shows that enhanced slip promotes earlier tear-film rupture. Physically, a higher slip coefficient reduces the viscous resistance to flow along the corneal surface. As a result, the surface velocities increase in both the $x$ and $z$-directions. This accelerates fluid motion from the valleys toward the crests leading to early tear film rupture. It is further evident from FIG \ref{fig:10}$(a)$ that partial slip reduces the rupture time for both smooth and rough substrates. Moreover, roughness amplifies the destabilising effect. FIG \ref{fig:10}$(b)$ shows the dependence of the tear breakup time $t_\text{rup}$ on the reduced capillary number $C$. The rupture time increases linearly with $C$, indicating that stronger capillary effects delay tear-film breakup. Since $C$ measures the relative influence of surface tension compared to viscous effects, larger values correspond to enhanced capillary stabilisation. Surface tension suppresses interfacial curvature and counteracts the destabilising van der Waals attraction, thereby reducing the rate of local thinning. As a result, the film remains stable for longer duration and rupture is delayed.\\
\begin{figure}
  \centering
  \includegraphics[width=0.7\textwidth]{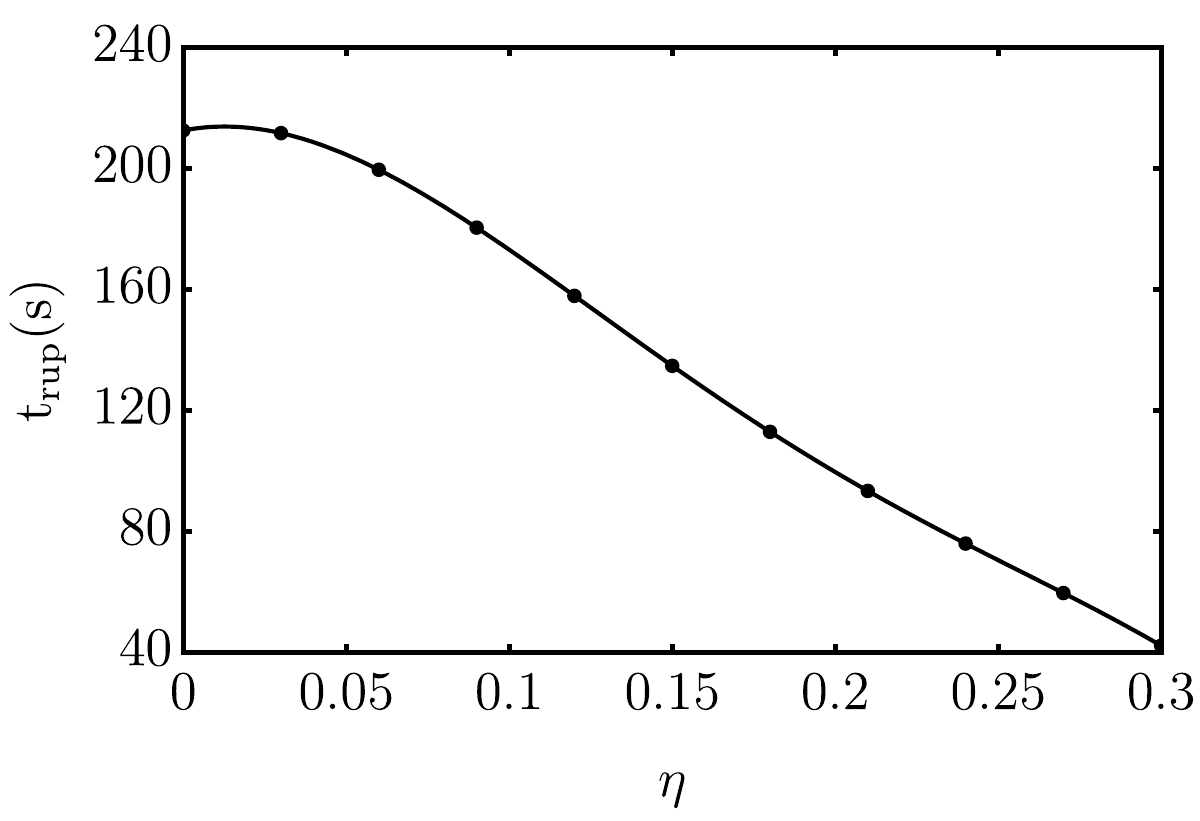}
  \caption{ Dependence of the dimensional tear breakup rupture time on $\eta$. The other parameters are $C=1,M=0.1,Pe_s=100,\beta=0.1,A_k=1$ and $\gamma_s=0.5.$ }
  \label{fig:11}
\end{figure}

FIG \ref{fig:11} illustrates the influence of the surface roughness amplitude $(\eta)$ on the dimensional tear rupture time $t_\text{rup}$ (s). As the amplitude of the roughness increases, $t_\text{rup}$ decreases. For very small $\eta$, the wall acts as a smooth surface. Hence, the rupture time remains nearly constant. However, beyond very small-amplitude roughness, $t_\text{rup}$ decreases significantly with increasing $\eta$. This indicates that higher surface irregularity enhances the destabilising mechanisms for film thinning. As a result, the elevated regions promote intensified thinning. Hence, the rupture occurs preferentially above these elevated regions. Over the range $ 0 \leq \eta \leq 0.3$, the predicted rupture time varies between approximately 214 s and 42 s. This clearly shows the strong sensitivity of tear-film stability to surface topography.\\

\begin{figure}
  \centering
  \includegraphics[width=1\textwidth]{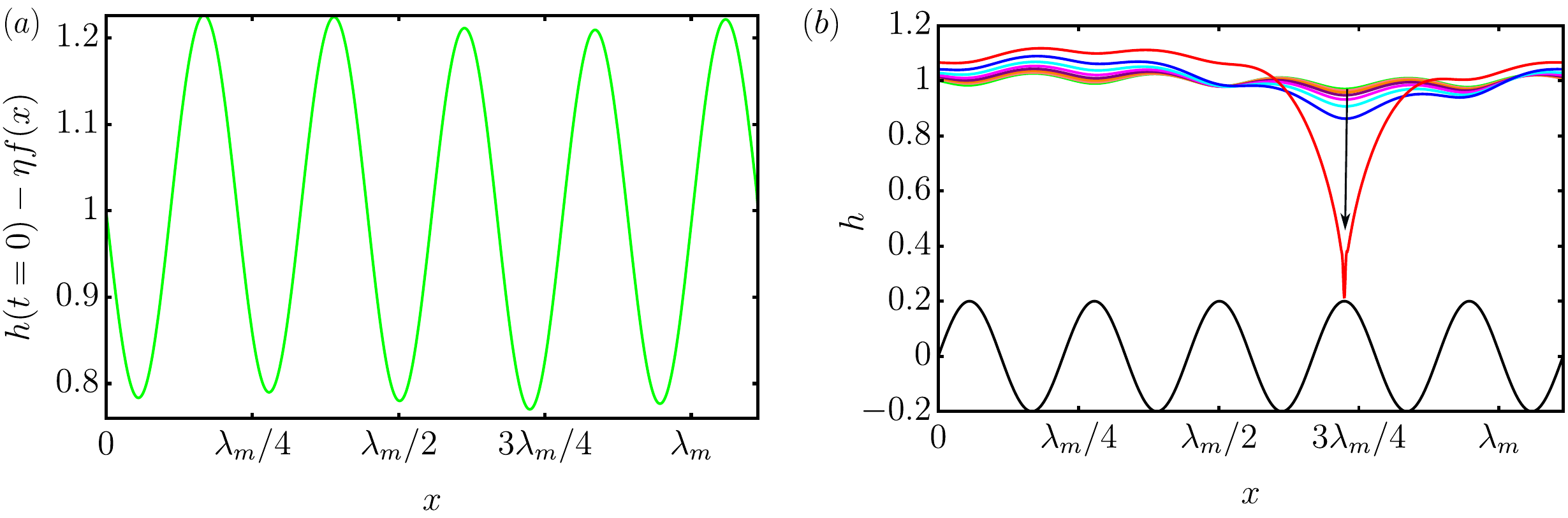}
  \caption{$(a)$ Effective Initial tear film thickness  $(b)$ temporal evolution of the tear film profile over a rough corneal surface  with ($\eta=0.2$) from numerical solution of equations (\ref{eq28}-\ref{eq29}). The initial condition is $h=h_s(x) +H_0 \sin (q_m x)$. The arrows indicate the progression of time from $t=0$ to the rupture time $t_\text{rup}$. The other parameters are $C=1,M=0.1,Pe_s=100,\beta=0.1,A_k=1$ and $\gamma_s=0.5, \eta =0.2$ } 
  \label{fig:12}
\end{figure}

We now examine the influence of the disturbance i.e., initial film profile on tear film rupture. FIG \ref{fig:12}$(a)$ shows effective initial tear film thickness $(h(t=0,x)-\eta f(x))$ over the rough corneal surface when the initial condition is a sinusoidal profile given by $h =h_s+H_0  \sin (q_m x)$. For this profile, the initial film thickness is minimum at $x=3.25 < 3 \lambda_m/4$ as shown in FIG \ref{fig:12}$(a)$. FIG \ref{fig:12}$(b)$ illustrates the corresponding nonlinear evolution of the film over the time. As the film thins, the van der Waals attraction from the elevated wall becomes increasingly dominant. As a result, the film is drawn toward this region leading to rupture at a location $x=3.25$ near $x=3 \lambda_m/4$.  In this case, rupture occurs at $t_\text{rup}=2.01$. However, for identical parameters but with an initial condition given by $h_s + H_0 \cos(q_m x)$, the rupture occurs near $x=\lambda_m/2$ as shown in FIG \ref{fig:6}$(a)$. This demonstrates that the rupture location is sensitive to the initial perturbation. In other words, the interaction between substrate roughness and the imposed disturbance determines where thinning is most strongly amplified.
\subsection{Comparison with the clinical data}
A meaningful assessment of the present predictions requires comparison with experimentally measured tear breakup times. Clinically, breakup time is measured using both invasive and non-invasive methods. Invasive techniques include fluorescein-based imaging whereas keratometry, retroillumination, and corneal topography measurements are non-invasive methods. Reported breakup times vary widely across studies. This variation is due to differences in measurement protocols. It is also influenced by physiological and environmental factors such as humidity, temperature, tear composition, blink pattern, and lighting conditions. In invasive methods, the concentration and chemical properties of fluorescein dye can also affect the measurements. As a result, reported tear breakup times for healthy eyes span a broad range in the literature as shown in TABLE \ref{tab:2}.\\
\begin{table}
\centering
\begin{tabular}{ll}
\hline
\textbf{Range of $t_{\mathrm{rup}}$ (s)} & \quad \textbf{Sources} \\
\hline
3-132 & \quad \cite{norn1969desiccation} \\
15-34 & \quad \cite{lemp1973factors} \\
5-100 & \quad \cite{vanley1977interpretation} \\
20-50 & \quad \cite{holly1977tear} \\
4-214 & \quad \cite{cho1992tear,korb2001comparison} \\
$>10$ & \quad \cite{sullivan2012clinical} \\
\hline
\end{tabular}
\caption{Reported ranges of tear film rupture time $t_{\mathrm{rup}}$ from various studies.}
\label{tab:2}
\end{table}
TABLE \ref{tab:2} summarizes the experimentally reported tear breakup times $(t_\text{rup})$ for healthy eyes. The reported values span a broad range, from as low as 3 s \citep{norn1969desiccation} to as high as 214 s \citep{cho1992tear,mengher1985non}. Several studies report typical rupture time between 15 and 50 s \citep{holly1977tear,lemp1973factors}.
The dimensional rupture times predicted by the present model lie in the range 42-212 s as shown in FIG \ref{fig:10}. These values lie in the higher range of the experimentally reported spectrum in TABLE \ref{tab:2}. These are consistent with measurements obtained under controlled environmental conditions. Notably, the predicted rupture time extend to several hundreds of seconds which align closely with the upper experimental range. Our results are also comparable with the rupture times reported  for a continuous mucin profile \citep{dey2019model}. In that study, the rupture times ranged from approximately 8-760 s for reduced capillary numbers in the range $10^{-2}$-$1$. While rupture time approaching 760 s is atypically large in clinical measurements, such values likely correspond to idealised smooth-surface conditions. Incorporating substrate roughness, as in the present model may yields rupture times that fall within a more physiologically representative range. Overall, this improved agreement supports our central argument that incorporating surface irregularity provides a more realistic representation of tear-film dynamics than models based on a perfectly smooth substrate. By including corneal roughness, the model captures destabilisation mechanisms that are likely present in vivo but absent in smooth-layer formulations.

\section{Conclusions} 
\label{sec:conclusions}
We investigate tear film rupture over a rough corneal surface driven primarily by van der Waals forces. The study is motivated by experimental observations indicating that the corneal surface is not perfectly smooth but exhibits microscopic protrusions and surface irregularities \citep{gipson2003role,king2014tear}. This study demonstrates that tear film dynamics over a rough corneal surface differ fundamentally from those predicted using idealized smooth-wall models.\\
The linear stability analysis based on Floquet theory revealed that the presence of wall roughness enhances instability in the tear film. The roughness affects the most unstable wave number, cutoff wavenumber and the growth rate of perturbation. This is validated by nonlinear simulations which demonstrated that wall roughness strongly influences the spatio-temporal evolution of the tear film. Surface roughness enhances local van der Waals attractions in regions of lower effective film thickness. This causes accelerated thinning and shifts the rupture site to the elevated site. Surface roughness reduces rupture time and the rupture location depends on the initial disturbance. The predicted dimensional rupture times (42-212 s) fall within the experimentally reported range for healthy tear films and align with previous theoretical studies, supporting the physiological relevance of the model.\\
These findings may help explain why tear film behavior in vivo often deviates from classical theoretical predictions and why ocular discomfort and contact lens complications persist despite advances in modelling and clinical practice. By incorporating physiologically realistic features such as corneal roughness and partial-slip conditions, this work offers a more complete representation of tear film instability. The results show that tear film rupture occurs earlier when the amplitude of roughness increases. They also show that a higher slip coefficient promotes earlier breakup. Both of these conditions are commonly associated with ocular surface disease and infection. This suggests that therapeutic approaches should not only target biochemical stabilization of the lipid layer but also prioritize restoring or preserving corneal smoothness to enhance tear film stability.\\
Future extensions of this model should incorporate additional physiological mechanisms such as evaporation, non-Newtonian rheology, and blink-induced shear flows. Including these effects will enable a more comprehensive understanding of tear film rupture under realistic conditions and may provide further insight into clinical strategies for managing dry eye and related disorders.

\begin{acknowledgments}
The authors acknowledge the Indian Institute of Technology Madras for providing research facilities. This work was supported by the Prime Minister’s Research Fellowship (PMRF). Deepak Kumar gratefully acknowledges Dr. Rajagopal Vellingiri for suggesting Fourier spectral method for the numerical simulations.
\end{acknowledgments}

\appendix

\section{Periodic coefficients in linear stability analysis}
\label{App:A}
The periodic coefficients are given as 

\begin{equation}
\begin{aligned}
P_1
=
\frac{4A_k}{\left(-\eta f(x)+h_s(x)\right)^4}
\Bigg[
(\eta f(x)-h_s(x))(3\beta-\eta f(x)+h_s(x))  
\times (\eta f''(x)-h_s''(x)) \\
+ \eta^2 (f'(x))^2 (-9\beta+2\eta f(x)-2h_s(x))  
+ 2\eta f'(x) h_s'(x) (9\beta-2\eta f(x)+2h_s(x))  \\
+ (h_s'(x))^2 (-9\beta+2\eta f(x)-2h_s(x))
\Bigg]
\end{aligned}
\end{equation}
\begin{equation}
Q_1
=
\frac{
A_k \left( 18\beta - 5\eta f(x) + 5h_s(x) \right)
\left( \eta f'(x) - h_s'(x) \right)
}
{\left( \eta f(x) - h_s(x) \right)^3}
\end{equation}
\begin{equation}
R_1
=
-\frac{
A_k \left( 3\beta - \eta f(x) + h_s(x) \right)
}
{\left( h_s(x) - \eta f(x) \right)^2}
\end{equation}
\begin{equation}
S_1
=
C \, (\eta f(x) - h_s(x))
(-2\beta + \eta f(x) - h_s(x))
(\eta f'(x) - h_s'(x))
\end{equation}

\begin{equation}
T_1
=
-\frac{1}{3} C
\left( h_s(x) - \eta f(x) \right)^2
\left( 3\beta - \eta f(x) + h_s(x) \right)
\end{equation}

\begin{equation}
U_1
=
- M
\left( \beta - \eta f(x) + h_s(x) \right)
\left( \eta f'(x) - h_s'(x) \right)
\end{equation}
\begin{equation}
V_1
=
\frac{1}{2} M
(\eta f(x) - h_s(x))
(-2\beta + \eta f(x) - h_s(x))
\end{equation}
\begin{equation}
\begin{aligned}
P_2
=
\frac{6 A_k \gamma_s}
{\left( \eta f(x) - h_s(x) \right)^5}
\Bigg[
(\eta f(x) - h_s(x))
(-2\beta + \eta f(x) - h_s(x))  
\times (\eta f''(x) - h_s''(x))
\\+ \eta^2 (f'(x))^2 (8\beta - 3\eta f(x) + 3h_s(x))
+ 2\eta f'(x) h_s'(x)
(-8\beta + 3\eta f(x) - 3h_s(x))  
\\+ (h_s'(x))^2
(8\beta - 3\eta f(x) + 3h_s(x))
\Bigg]
\end{aligned}
\end{equation}
\begin{equation}
Q_2
=
-\frac{
3 A_k \gamma_s
\left( 7\beta - 3\eta f(x) + 3h_s(x) \right)
\left( \eta f'(x) - h_s'(x) \right)
}
{\left( h_s(x) - \eta f(x) \right)^4}
\end{equation}
\begin{equation}
R_2
=
\frac{
3 A_k \gamma_s
\left( 2\beta - \eta f(x) + h_s(x) \right)
}
{2 \left( \eta f(x) - h_s(x) \right)^3}
\end{equation}
\begin{equation}
S_2
=
C \gamma_s
\left( \beta - \eta f(x) + h_s(x) \right)
\left( \eta f'(x) - h_s'(x) \right)
\end{equation}
\begin{equation}
T_2
=
\frac{1}{2} \, C \gamma_s
\left( \eta f(x) - h_s(x) \right)
\left( 2\beta - \eta f(x) + h_s(x) \right)
\end{equation}
\begin{equation}
U_2
=
M \gamma_s
\left( h_s'(x) - \eta f'(x) \right)
\end{equation}
\begin{equation}
V_2
=
M \gamma_s
\left( \beta - \eta f(x) + h_s(x) \right)
+
\frac{1}{Pe}
\end{equation}

\section{Derivation for purely imaginary Floquet exponents}
\label{App:B}
The solution $h_1$ and $\gamma_1$ are required to satisfy periodic boundary conditions on the domain $[0,1]$. We therefore employ a Floquet representation and assume perturbations of the form, $h_1 (x)=e^{\sigma t} e^{\alpha x} \phi(x)$ and $\gamma_1 (x)=e^{\sigma t} e^{\alpha x} \psi(x)$. Here, the Floquet exponent $(\alpha)$ is a complex number. $\phi(x)$  and $\psi(x)$ are periodic with period 1. Imposing periodicity of the perturbations gives $\phi(x+1)=\phi(x)$ and $\psi(x+1)=\psi(x)$. Substituting the Floquet form gives
\begin{equation}
    h_1 (x+1)= e^{\sigma t}e^{\alpha (x+1)}  \phi(x+1)=e^{\sigma t}e^\alpha e^{\alpha x} \phi(x)
\end{equation}
and
\begin{equation}
    \gamma_1 (x+1)=e^{\sigma t}e^{\alpha (x+1)}  \psi(x+1)=e^{\sigma t}e^\alpha e^{\alpha x} \psi(x)
\end{equation}
Hence, periodicity demands $e^\alpha=1$. Therefore, $\alpha=2 \pi in$ where $n \in \mathbb{Z}$ and thus the Floquet exponent $(\alpha)$ is purely imaginary. 
\nocite{*}

\bibliography{apssamp}

\end{document}